\documentclass[11pt,a4paper]{article}
\pdfoutput=1
\usepackage{jcappub}

\usepackage{amsmath, amsthm, amssymb, amsfonts}
\usepackage{slashed}
\usepackage[ngerman,english]{babel}
\usepackage{epsfig, graphicx}
\usepackage{color}
\usepackage{afterpage}

\newcommand{\U}{\ensuremath{\text{U(1)}_\text{h}}}

\newcommand{\dd}{\text{d}}

\newcommand{\be}{\begin{equation}}
\newcommand{\ee}{\end{equation}}
\newcommand{\bea}{\begin{equation}\begin{aligned}}
\newcommand{\eea}{\end{aligned}\end{equation}}
\newcommand{\bse}{\begin{subequations}}
\newcommand{\ese}{\end{subequations}}

\newcommand{\neff}{\ensuremath{N_\text{eff}} }

\newcommand{\TDS}{\ensuremath{T_{\text{DS}}} }
\newcommand{\Tnu}{\ensuremath{T_{\nu}}}
\newcommand{\TG}{\ensuremath{T_{\gamma}}}

\newcommand{\W}{\mathcal{W}}

\begin{document}

\subheader{\hfill MPP-2013-224}

\title{Dark Radiation constraints on minicharged particles in models with a hidden photon}

\author[a]{Hendrik~Vogel }
\author[a,b]{, Javier~Redondo}

\affiliation[a]{Max-Planck-Institut f\"ur Physik, 
F\"ohringer Ring 6, D-80805 M\"unchen, Germany}
\affiliation[b]{Arnold Sommerfeld Center, 
Ludwig-Maximilians-Universit\"at, Theresienstr.~37, D-80333 M\"unchen, Germany}

\emailAdd{hvogel@mpp.mpg.de}
\emailAdd{redondo@mpp.mpg.de}

\abstract{

We compute the thermalization of a hidden sector consisting of minicharged fermions (MCPs) and massless hidden photons in the early Universe. The precise measurement of the anisotropies of the cosmic microwave background (CMB)  by Planck and the relic abundance of light nuclei produced during big bang nucleosynthesis (BBN) constrain the amount of dark radiation of this hidden sector through the effective number of neutrino species, \neff. This study presents novel and accurate predictions of dark radiation in the strongly and weakly coupled regime for a wide range of model parameters. We give the value of \neff for MCP masses between $\sim$ 100 keV and $10$ GeV and minicharges in the range $10^{-11}-1$. 
Our results can be used to constrain MCPs with the current data and they are also a valuable indicator for future experimental searches, should the hint for dark radiation manifest itself in the next release of Planck's data.

}

\maketitle

\section{Introduction}

Extensions of the Standard Model (SM) of particle physics with an additional \emph{hidden} $\U$ gauge symmetry have recently gathered a wealth of attention. In the simplest realization~\cite{Galison:1983pa,Holdom:1986eq}, the only new particle included is a gauge boson which has received many names: paraphoton~\cite{Okun:1982xi},  hidden photon~\cite{Ahlers:2007qf} or dark photon~\cite{Bjorken:2009mm} to name but a few. 
Despite the fact that the particles of the SM are all singlets under the new $\U$ (hence it being \emph{hidden}), the hidden photon (HP henceforth) can have kinetic mixing with the hypercharge boson. It is radiatively generated if there exist ``mediator'' fields (irrespective of how large their mass is) charged under both gauge groups~\cite{Galison:1983pa,Holdom:1985ag}. 
The most natural value of the kinetic mixing parameter is thus $\chi \sim  g' \times \mathcal{O}(10^{-3})$ with $g'$ the hidden gauge coupling. Much smaller values occur when $g'$ is very small, like in~\cite{Antoniadis:2001np,Burgess:2008ri,Goodsell:2009xc}, or when cancellations among different mediators happen, for instance if any of the U(1)'s is embedded in a non-abelian group. Typical values predicted in the literature yield a range $\chi\sim 10^{-12}-10^{-3}$~\cite{Dienes:1996zr,Lukas:1999nh,Abel:2003ue,Blumenhagen:2005ga,Abel:2006qt,Abel:2008ai,Goodsell:2009pi,Goodsell:2009xc,Bullimore:2010aj,Goodsell:2010ie,Cicoli:2011yh,Heckman:2010fh,Goodsell:2011wn}.

If the HP obtains a mass via the St\"uckelberg or Higgs mechanisms, it can be produced in any reaction among SM particles in which an ordinary photon (or Z boson) is produced. 
Many laboratory experiments at the low-energy frontier are testing the existence of these particles: 
accelerator based searches~\cite{Aubert:2009cp,Jaeckel:2012yz,Gninenko:2013sr}, beam dump~\cite{Bjorken:2009mm,Batell:2009di,Essig:2010xa,Freytsis:2009bh,Wojtsekhowski:2009vz,Merkel:2012ce,Andreas:2012mt}, helioscopes~\cite{Redondo:2008aa, Schwarz:2011gu,Mizumoto:2013jy,An:2013yua,Redondo:2013lna}, laser and microwave photon regeneration experiments~\cite{Redondo:2010dp,Ehret:2010mh,Bahre:2013ywa,Jaeckel:2007ch,Slocum:2010zz,Williams:2010zz,Povey:2011ak,Betz:2012ce} and searches for 
spectral features in light propagation over astronomical distances~\cite{Mirizzi:2009iz,Lobanov:2012pt}, 
see~\cite{Hewett:2012ns,Jaeckel:2013ija} for recent reviews. 
HPs are also thermally produced in the early Universe, the relic population behaving as dark matter~\cite{Pospelov:2008jk,Redondo:2008ec} or dark radiation~\cite{Jaeckel:2008fi} depending on its mass (100 keV or meV respectively)\footnote{Dark matter HPs can have their origin in the misalignment mechanism, in which case their mass can have much broader values~\cite{Nelson:2011sf,Arias:2012az}. }.  

If the HP is massless, it has no phenomenological consequences whatsoever because the probability of producing a quantum of HP is proportional to the HP's mass.  
At the Lagrangian level, the only difference from a pure SM is a small renormalization of the hypercharge gauge coupling, $g=g(\chi)$. 
Since in practice $g$ has to be measured one cannot know whether it contains a hidden contribution or not. However, if the hypercharge U(1) unifies with SU(2) weak isospin in a grand unified model, the value of $g$ can be calculated theoretically given the weak coupling and one can constrain a HP contribution~\cite{Redondo:2008zf}. 

The unbroken hidden $\U$ case becomes very interesting when we consider additional particles with hidden $\U$ charge. Because of the small $\chi$ mentioned above, these particles appear as if they had a small electric charge $\epsilon=g'\chi/e$~\cite{Holdom:1985ag} and we call them \emph{minicharged particles} (MCPs henceforth)\footnote{MCPs appear in different constructions in extensions of the standard model, see also~\cite{Batell:2005wa,Bruemmer:2009ky}.}.  
The existence of this type of MCPs does not challenge the standard arguments of the existence of magnetic  monopoles and  the quantization of charge but makes them more subtle~\cite{Brummer:2009cs}. 

Since the pioneer works~\cite{Galison:1983pa,Holdom:1985ag} many experiments and phenomenological arguments have been devised to put the existence of MCPs into test. 
Direct laboratory searches for MCPs have been performed in accelerators~\cite{Davidson:1991si}, 
a dedicated beam dump experiment at SLAC~\cite{Prinz:1998ua} and ortho-positronium decays~\cite{Badertscher:2006fm}.    
For MCPs of low mass ($m_f<30$ keV) the most relevant constraints come from stellar evolution. 
The stellar energy loss due to the emission of MCP pairs by plasmon decay has a number of consequences that can be constrained~\cite{Davidson:1991si,Raffelt:1996wa}. It delays the helium flash in red-giants (brightening the tip of the red-giant branch), accelerates the helium-burning stage~\cite{Davidson:1991si,Davidson:1993sj,Davidson:2000hf} and the cooling of white-dwarves~\cite{Davidson:1991si,Davidson:2000hf} and would have reduced the neutrino pulse of SN1987A~\cite{Mohapatra:1990vq,Davidson:1993sj}. 

In this paper we focus on cosmological probes of \emph{minicharged particles in models with a massless hidden photon}. 
MCPs created in the early Universe can behave as dark matter (DM) and/or dark radiation (DR) and the HPs contribute to DR. 
Current cosmological data severely constrains the amount of DM, its possible interactions with the baryon+photon fluid and with itself, and the amount of DR. 
In order to translate this into bounds on the MCPs' and HPs' parameters one needs to accurately calculate the 
production and decoupling of MCPs and HPs in the early Universe. In the pioneer work of Davidson, Campbell and 
Bailey~\cite{Davidson:1991si} these calculations were done analytically in simple approximations.
They presented two important cosmological bounds. First, they derived an overclosure bound from requiring the relic density of MCP DM to be smaller than the critical density today ($\Omega<1$). 
Second, they used the contemporary constraint on dark radiation~\cite{Steigman:1986nh} (traditionally expressed in terms of the effective number of neutrino species $N_{\rm eff}$) from the helium-4 produced in big bang nucleosynthesis (BBN). In the early Universe, MCPs and HPs created during reheating or by interactions with the SM thermal bath quickly come into thermal equilibrium with each other (we assume that $g'$ is not hyper weak) constituting a thermal ``dark sector'' (DS henceforth).  
If the kinetic mixing is large enough, the DS and the SM thermalize with each other as well. 
If these DS particles are relativistic during BBN, their contribution to $N_{\rm eff}$ is $2+8/7=3.14$ (for a Dirac fermion MCP), which was ruled out by data back then.  
MCPs with masses $m_f\gtrsim$ MeV can avoid this bound. 
When the Universe's temperature reaches the MCP mass, the thermal abundance of MCPs becomes exponentially suppressed and, from that moment on, HPs have no means of interacting with the SM and  decouple. 
All SM particles that become non-relativistic afterwards give their entropy to the SM bath heating it with respect to the DS.  
If $m_f$ is sufficiently above MeV (the key temperature range for the BBN bound) there are enough particle species in the SM to dilute the HP density below the observational bound.  
This lead to the very strong bound: $m_f>200$ MeV. 
Alternatively the MCPs should have never been in thermal equilibrium with the SM bath, which was estimated to happen for $\epsilon<10^{-8}$. 
However, in that work it was incorrectly assumed that MCPs give all their entropy to HPs when they decouple at $T\sim m_f$.  
Since the amount of HPs was overestimated, so was the lower limit on the mass. 

In a later paper~\cite{Davidson:1993sj}, the limit was corrected to $m_f>$ MeV after observing that for $\epsilon>10^{-8}$  the MCPs annihilate while still being in thermal equilibrium with the SM. 
Their energy is thus split between HPs and SM particles reducing the value of $\neff$.
The situation for couplings around $\epsilon\lesssim 10^{-8}$ was never considered in any detail. 
It realizes an intermediate case between the assumption of all MCP entropy going to HPs (assumed in \cite{Davidson:1991si}) and being distributed equally into the SM and HP thermal populations. 
In this range of parameters one expects the bounds to strengthen because the DS comes close to equilibrium with the SM but the coupling between the SM and DS can be weak enough to favor the MCP entropy flow into the HPs, enhancing $N_{\rm eff}$.      

The purpose of this paper is to update the cosmological constraints from dark radiation on MCPs in models with a HP,  
treating production and decoupling in full glory.  
This is timely because of the interest raised on these particles and hidden sectors in general and the considerable amount of cosmological data made available in the last decade.  
The nuclear reaction network of BBN is now better understood and brand new data on primordial element abundances has been collected and analyzed (especially deuterium~\cite{Pettini:2012ph,Cooke:2013cba} and helium~\cite{Izotov:2010ca,Izotov:2013waa}). 
The upper limit on the helium-4 abundance $Y_p<0.2631$ (95\% C.L.)~\cite{Mangano:2011ar} stands as 
a reliable figure, regardless of assumptions on stellar processing or the uncertainties on the primordial baryon density and the neutron lifetime.  

Furthermore, nowadays we have complementary information on the amount of dark radiation provided by the temperature anisotropies of the cosmic microwave background (CMB). 
Only recently the WMAP mission achieved enough precision to assess the existence of a cosmic neutrino background~\cite{Dunkley:2008ie}, i.e. $\neff>0$. Combining CMB data with other late cosmology data sets -- large scale structure, baryon acoustic oscillations (BAO) and direct measurements of the Hubble constant by the Hubble space telescope (HST) -- improves the measurements of $N_{\rm eff}$. The obtained values tended to be larger than 3, raising the excitement of a possible hint of new physics, see e.g.~\cite{Archidiacono:2013lva,Abazajian:2012ys,Calabrese:2013jyk}. The latest CMB results of WMAP, SPT~\cite{Keisler:2011aw} and ACT~\cite{Dunkley:2010ge} combined with BAO and HST gave $\neff= 3.84\pm 0.40$ at $68\%$ C.L.~\cite{Hinshaw:2012aka}. Thus, although somehow controversial~\cite{Feeney:2013wp}, rather than a constraint there seemed to be a 2-$\sigma$ preference for a non-negligible amount of unaccounted dark radiation~\cite{Hamann:2011hu}.  
The recent results of the Planck mission~\cite{Ade:2013zuv} have unfortunately not clarified the issue.  
Combined with WMAP polarization (WP) maps, SPT and ACT, BAO and HST the Planck teams gives $N_{\rm eff}=3.52^{+0.48}_{-0.45}$ at 95\% C.L.~\cite{Ade:2013zuv}. 
Although the error in $\neff$ has decreased according to the expectations~\cite{Hamann:2007sb}, the central value has done so too in such a way that the 2-$\sigma$ excess remains. 
The Planck analysis have brought more information, revealing an increased tension between the HST direct measurement of the Hubble constant $H_0=73.8\pm 2.4$ km/(Mpc s)~\cite{Riess:2011yx} and the lower estimate $H_0=67.3\pm 1.2$ km/(Mpc s) using Planck and other CMB data alone~\cite{Ade:2013zuv}. 
The value of $H_0$ is positively correlated with $\neff$ in cosmological fits (see~\cite{Hou:2011ec} and figure 21 of~\cite{Ade:2013zuv}) so that using the HST prior tends to push $\neff$ to higher values. 
In other words, a high $\neff$ softens the tension between CMB and local probes of $H_0$ but a 
systematic bias of the local measurements towards high-$H_0$ could be artificially triggering the excess\footnote{A discussed alternative, that our visible Universe is placed in a local underdensity, can only relieve a small part of the tension~\cite{Marra:2013rba}.}. It is also worth mentioning that although $\neff>3$ reduces the tension of $H_0$,  it worsens the agreement between CMB and local measurements of the age of the Universe~\cite{Verde:2013wza}. This reduction is in any case not significant~\cite{Verde:2013cqa,Verde:2013wza,Ade:2013zuv}. 
When the HST prior is excluded from the analysis, the Planck team finds $N_{\rm eff}=3.30^{+0.54}_{-0.52}$ at 95\% C.L.~\cite{Ade:2013zuv} and this is the value that we will use in this work. 
Planck has the potential to improve its sensitivity to $\neff$ down to the $\pm 0.2$ level~\cite{Hamann:2007sb} and future polarization measurements can decrease this figure to the $0.05$ level~\cite{Galli:2010it}. 
Thus there is still hope for a significant detection of DR in the future. We shall then present our results in a flexible 
and detailed way to allow the future user to derive stronger constraints or identify MCP parameters that fit an excess.  

With this target in mind we have computed in detail all the processes leading to the production and decoupling 
of MCPs and HPs  in the early Universe to track the amount of dark radiation present during BBN and later on 
during the CMB epoch. 
We can track the evolution of the energy density in the DS even for parameters where its  
coupling to the SM is only mild and thermalization with the SM bath is never complete. 
Pertaining this, we acknowledge a very comprehensive study of dark radiation in general extensions of the SM~\cite{Brust:2013ova} which appeared recently. This excellent work covers partially the scope of this paper, touching on the MCP+HP case (sec 3.3.1~\cite{Brust:2013ova}). In comparison, we focus \emph{exclusively} on it so we can explore a wider parameter space of couplings and masses, and discuss the role of the different production and decoupling channels. However, our works are complementary because~\cite{Brust:2013ova} considers a range of different initial temperatures of the dark sector while we set it to zero to obtain conservative constraints. 

\begin{figure}[t] \centering
 \includegraphics[width=0.7\textwidth]{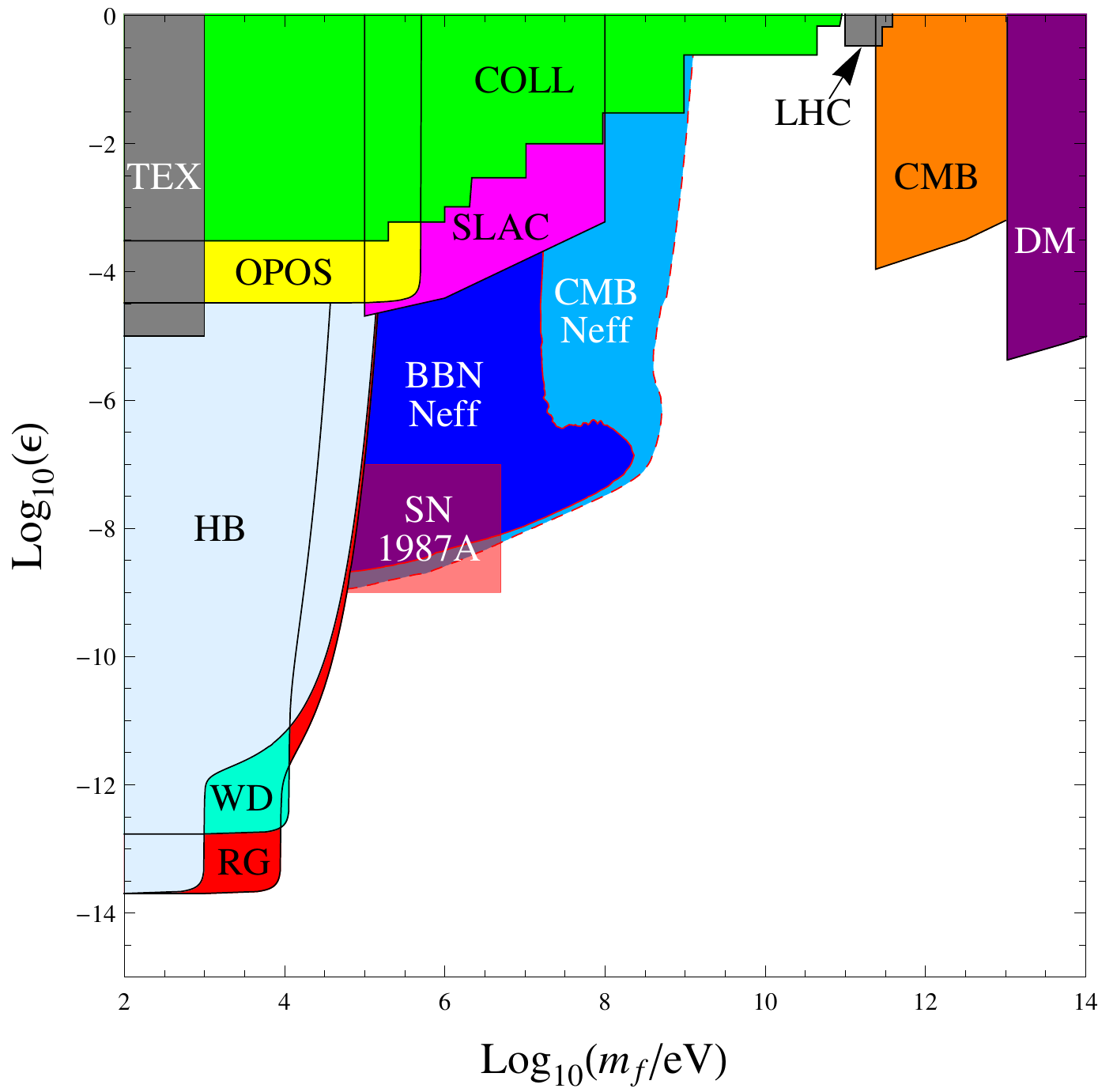}	
 \caption[Exclusion plot for MCPs]{{\small{
Summary of constraints on fermionic MCPs in the mass/minicharge plane for $g'=0.1$. 
The results of this work are: the constraint on \neff during BBN (dark blue) and on \neff by Planck (light blue). 
We have also improved the bounds from white dwarves (WD), red giants (RG) and horizontal branch (HB) with respect to the originals by calculating the high mass behavior. 
The remaining bounds are taken from elsewhere: LHC~\cite{Jaeckel:2012yz}, DM~\cite{Davidson:1991si}, COLL~\cite{Davidson:1991si}, SLAC~\cite{Prinz:1998ua}, OPOS~\cite{Badertscher:2006fm}, TEX~\cite{Gninenko:2006fi} and CMB~\cite{Dubovsky:2003yn,Dolgov:2013una} (see also appendix~\ref{CMBupdate}). }}} \label{fig:MCPresult} \end{figure}

Our results are summarized in figure~\ref{fig:MCPresult} where we also show the most relevant constraints on MCPs in models where the minicharge arises as a consequence of kinetic mixing ($g'=0.1$). Other interesting constraints which are not shown have been discussed in~\cite{Ahlers:2009kh,Gies:2006hv,Jaeckel:2009dh,Melchiorri:2007sq,Gluck:2007ia,Ehret:2010mh,Burrage:2009yz}. 
The Planck $\neff$ constraint  disfavors MCPs with $m_f<$ GeV down to $\epsilon\sim 10^{-8}-10^{-7}$  but leave a small region around $m_f\sim 5$  MeV (to be discussed further on). The BBN constraints cover this gap. They are similar to previous ones except for the region $m_f\sim 100$ MeV and $\epsilon\sim 10^{-7}$ where we find the mentioned strengthening of the BBN constraint due to the weak coupling between the hidden and SM sectors. 
Since our constraints have some overlap with astrophysical bounds at the lowest masses we computed the high-mass boundary of the helium-burning (HB), red-giant (RG) and white dwarf (WD) bounds more accurately. We also included the recent update~\cite{Dolgov:2013una} on MCPs' acoustic oscillations during recombination~\cite{Dubovsky:2003yn}.

The plan of the paper is as follows. In sections 2 and 3 we present our definitions of the MCP+HP 
Lagrangian extending the SM. In section 4 we describe the equations and reactions ruling the evolution of the 
energy density of the hidden sector. In section 5 We present the bounds coming from 
$N_{\rm eff}$ at the CMB epoch and explain different examples of the thermal histories  
encountered in different regions of parameter space. In section 6 we focus on the constraints from BBN and in section 7 we present our conclusions. The revision of the astrophysical bounds at high masses and the update on MCPs' acoustic oscillations is done in the appendix. 

\section{The model}

In this article we extend the SM gauge group by an additional unbroken local \U under which all SM particles are singlets. We also add a massive hidden fermion charged under the new \U \ only. The additional terms to the SM Lagrangian then read
\be
\mathcal{L} = -\frac{1}{4}F'_{\mu \nu}F'^{\mu \nu} + \bar f (i\slashed{D}-m_f) f -\frac{\chi}{2}F_{\mu \nu} F'^{\mu \nu},
\ee
where $F'_{\mu \nu}, F_{\mu \nu}$ are the field strength tensor of the HP and SM photon, respectively,  $f$ denotes the hidden fermion, $m_f$ its mass and $\chi$ is the kinetic mixing parameter after electroweak symmetry breaking. The covariant derivative is
\be
D_\mu=\partial_\mu -i g' A'_\mu,
\ee
where $g'$ is the gauge coupling of the \U\ and $A'_\mu$ is the vector potential of the HP.

Our results are only negligibly affected by physics at high energy scales so that we do not include mixing with the $Z^0$ boson or corrections from possible UV completions (e.g. SUSY).
\\
Since the \U\ is unbroken, the HP remains massless. The kinetic part of the Lagrangian can be brought into the canonical form by rotating away the kinetic mixing through the redefinition $A'_\mu\to A'_\mu-\chi A_\mu$. Without the coupling to $f$, the resulting $A'$ would be completely decoupled from the SM and thus unobservable. 
Including the hidden fermion, however, the redefinition induces a term $-g' \chi \bar f \slashed{A} f$, i.e. the fermion gets an electric charge $\epsilon = g' \chi/e$ ($e$ is the electron charge). 
 Since $\chi$ is typically small, so is the electric charge. Therefore, $f$ is called a minicharged particle (MCP).

The gauge coupling $g'$ can be of order unity and in the following we assume that the coupling between HPs and MCPs is strong enough to keep them in local thermal equilibrium (LTE) with a temperature \TDS at all times. 
We check the limits of this assumption in section~\ref{initialconditions}.

\section{Number of effective neutrinos}

The HPs and the MCPs contribute to the radiation density of the Universe. 
Assuming HPs and MCPs in thermal equilibrium the energy density of the dark sector can be computed in terms of a common dark sector temperature, $\TDS$,  and an effective number of DS relativistic degrees of freedom, $g_{*\text{DS}}$, as 
\be
\rho_{\rm DS} = \rho_{\rm HP}+  \rho_{\rm MCP}=\frac{\pi^2}{30}\TDS^4 g_{*\text{DS}}(z) = 
\frac{\pi^2}{30}\TDS^4 \left(2+ \frac{7}{2}\int_z^\infty\frac{\sqrt{x^2-z^2}x^2dx}{e^x+1}\frac{120}{7 \pi^4}\right)
\ee
where $z=m_f/\TDS$ and the integral is defined to be 1 for $z\to 0$. 
The HP is massless and always contributes, while the MCP contribution is exponentially suppressed once their temperature falls below their mass. The spectrum of the CMB is sensitive to the amount of radiation in the Universe at the epoch of matter-radiation equality and decoupling ($\TG\sim \mathcal{O}$(eV)), and Planck~\cite{Ade:2013zuv} was able to measure the energy density of radiation with unprecedented precision. 

The energy density of radiation is usually parametrized by the effective number of neutrinos $\neff$. 
At the CMB epoch this is defined through
\be
\rho_\text{R}=\frac{\pi^2}{30}\left[g_{\gamma}+2\frac{7}{8}\left(\frac{4}{11}\right)^{4/3}\neff \right]T_{\gamma}^4, \label{eq:neff}
\ee
where $g_\gamma=2$ are the photon's degrees of freedom. 
In standard cosmology with only the SM particle content, $\neff=3.046$ \cite{Mangano:2001iu}. In our model \neff includes the contribution of the HPs and MCPs as well. 
The total \neff then reads
\be
\neff = 3\left(\frac{11}{4}\right)^{4/3}\left(\frac{T_\nu}{\TG}\right)^4+\frac{8}{7}\left(\frac{11}{4}\right)^{4/3}\left[\frac{g_{*\text{DS}}(m_f,\TDS)}{2}\right]\left(\frac{\TDS}{\TG}\right)^{4}. \label{CMB formula}
\ee
where \TDS, $\Tnu$ and $\TG$ are the temperatures of the DS, neutrinos and photons, respectively. The first term is the neutrino contribution, which in the MCP scenario can significantly deviate from the standard value. 

\section{Equations for the SM-DS energy transfer}

To obtain \neff at the CMB epoch, we have to track the temperature ratios $\TDS/\TG$ and $\Tnu/\TG$ during the evolution of the Universe. 

For the temperature of the DS, we study the evolution of the SM and DS energy densities with time.  We use the following set of coupled differential equations

\bse\label{system1}
\begin{align}
&\dot \rho_{\text{SM}} + 3H\left(\rho_{\text{SM}}+P_{\text{SM}}\right)=-\W,\\
&\dot \rho_{\text{DS}} + 3H\left(\rho_{\text{DS}}+P_{\text{DS}}\right)=\W,
\end{align}
\ese
where $\rho_{\text{SM}} $ ($\rho_{\text{DS}}$) is the energy density of the SM (DS) particles, $P_{\text{SM}}$ ($P_{\text{DS}}$) is the pressure of the SM (DS) particles, $\dot{ }$ denotes a derivative with respect to time, and the source term $\W$ encodes the energy transported from the SM to the DS sector per unit time by particle reactions, described further in section \ref{source}. $H$ is the Hubble parameter,  
\be
H^2=\frac{8\pi }{3 M_p^2}\rho_{\text{total}}=\frac{8\pi}{3 M_p^2}\left(\rho_{\text{SM}}+\rho_{\text{DS}}\right), 
\ee
where $M_p=1.22\times 10^{19}$ GeV is the Planck mass. The neutrino energy density is contained in $\rho_\text{SM}$ up to $\TG\sim 3$ MeV. We instantly decouple neutrinos at that temperature, afterwards tracking their energy density separately. 

If the SM and DS are very strongly coupled (high $\epsilon$) we should have of course $\TG=\TDS$ and the 
cooling of the Universe is ruled by quasi-adiabatic expansion, in which the comoving entropy is conserved. In this case, one can compute \neff explicitly as a function of the SM-DS decoupling temperature with the formulas developed in~\cite{Blennow:2012de}. The decoupling temperature in this case is never far from $m_f$, when the MCP population gets exponentially suppressed. If we want to know the temperature more precisely, we need to accurately compute the time when the energy transfer between the sectors becomes inefficient by using the above equations \eqref{system1}. 
We have cross-checked our results with the entropy conservation hypothesis to find good agreement for large $\epsilon$ using a decoupling temperature $T\sim m_f/10$. For this comparison we have extended the formulas of~\cite{Blennow:2012de} to cover smoothly the cases considered there, see appendix~\ref{app1}.

\subsection{Source term \label{source}}

The source term $\W$ of equations \eqref{system1} is the particle physics' input on how efficiently the DS and the SM exchange energy. The most relevant reactions are 2 to 2 processes so their contribution can be written as a sum of terms of the sort

\bea
\W &=\int \dd \Pi_a \dd \Pi_b \dd \Pi_c \dd \Pi_d\  (2\pi)^4\delta^4(p_a+p_b-p_c-p_d)\\&\quad\quad
\times E_{\text{trans}} \times | \mathcal{\widetilde M}|^2_{a+b\rightarrow c+d}\left(f_af_b-f_cf_d\right),\label{eq:gamma}
\eea
where $| \mathcal{\widetilde M}|^2_{a+b\rightarrow c+d}$ is the matrix element for the reaction $a+b\to c+d$ summed over initial and final polarizations and $E_{\text{trans}}$ is the transported energy per collision. $f_x$ is the phase-space density, which for particles in LTE is either a Bose-Einstein or Fermi-Dirac distribution. To reduce the number of integrals analytically as in~\cite{Gondolo:1990dk} we approximate the distributions by Maxwell-Boltzman type. Note that we neglect blocking and stimulation factors whose effects are always small. 
The one-particle phase space differential volume is 
\be
\dd \Pi_x= g_x\frac{\dd^3 p_x}{(2 \pi)^3 2 E_x}
\ee
where $g_x$ denotes the internal degrees of freedom of particle $x$ besides spin, which is included in the matrix element squared. 

To leading order, the following channels contribute to the source term $\W$
\bse \label{eq:processesT}
\begin{align}
e^+e^- \leftrightarrow f\bar f  &, \quad W^+W^- \leftrightarrow f\bar f, \label{eq:SMannihilation}\\
\gamma^* &\leftrightarrow f \bar f,\label{plasmon decay}\\
f\bar f &\leftrightarrow \gamma \gamma' \label{eq:annihi},\\
f\bar f &\leftrightarrow \gamma \gamma \label{eq:annihi2},\\
\gamma f &\leftrightarrow \gamma' f,\label{eq:comptscat}\\
\gamma f &\leftrightarrow \gamma f,\label{eq:comptscat2}\\
e^-f &\leftrightarrow e^-f,\label{eq:coulomb}
\end{align}
\ese
where particles can be replaced by the corresponding antiparticles for scattering, and $e^+e^-$ can be replaced by other electrically charged particle/antiparticle pairs including mesons like $\pi^+\pi^-$.

We divide the processes into three classes: Those that are efficient when the population of DS particles is very small (``production channels''), those that are most efficient when the DS population is sizable (``decoupling channels''), and other channels, which give small corrections.

\subsubsection{Production channels \label{prod}}

SM particle pair-annihilation \eqref{eq:SMannihilation} and plasmon decay \eqref{plasmon decay} into MCPs are efficient even in the absence of a DS thermal bath. These channels produce an abundance of DS particles and bring the DS and the SM sector closer to equilibrium. 
The energy transfer normalized to the equilibrium value ($\sim \TG^4$) goes as $\W/\rho_{\rm SM}\simeq \TG^5/\TG^4$ when both species (the SM annihilated and DS created) are relativistic and decreases exponentially when one of their masses becomes smaller than $\TG$. The time interval is $dt= d\TG/H\TG\propto d\TG/\TG^3$ (radiation domination) and thus the integrated energy transferred, $\int \W dt/\TG^4\propto 1/\TG^{1}$, is dominated by the smallest  temperatures where both species are still relativistic, $\TG^1\sim {\rm max}\{m_f,m_{\rm SM}\}$. The contribution from a heavy SM particle is thus inversely proportional to its mass unless it is lighter than the MCP mass. Since we cannot probe MCP masses much above the GeV scale, we neglect contributions of $W^\pm$ and $t\bar t$ to the annihilation \eqref{eq:SMannihilation}. 

Before the QCD phase transition ($\Lambda_{\text{QCD}} \sim 180 \ \text{MeV}$~\cite{Wantz:2009it, Coleman:2003hs}) we include all contributions from elementary particles with masses smaller than the $W$-bosons. Afterwards, we should replace the contributions from quarks by mesons. 
As a compromise between simplicity and accurateness at the lowest energies we have only considered  the contribution from the charged pions.
Mesons and baryons more massive than the pions have their abundances already exponentially suppressed at the QCD phase transition already\footnote{The contribution of the $\pi^0$ is small which justifies to neglect it as well.}. 

For the pair production process~\eqref{eq:SMannihilation}, the matrix element in the center of mass frame is 
\be
|\widetilde{\mathcal{M}}|^2_{l^+l^-\rightarrow f+\bar f}=
4e^4 Q_l^2 \epsilon^2 \frac{s^2 (1+\cos^2{\theta})+4 s(m_f^2+m_l^2) (1-\cos^2{\theta})+16 m_f^2 m_l^2 \cos^2\theta}{s^2},
\label{eq:matrixffee}
\ee
where $s$ is the center of mass energy squared, $m_l$ the SM particle mass and $Q_l$ its electric charge. 
In practice, the lighter of the two masses has a subdominant effect on $\W$ so we neglect its contribution. 
In $\pi^+,\pi^-$ annihilation, we have a similar expression 
\be
|\widetilde{\mathcal{M}}|^2_{\pi^+\pi^-\rightarrow f+\bar f}=
2e^4  \epsilon^2 |F_\pi|^2 \frac{s^2 (1-\cos^2{\theta})+4 s[m_f^2\cos^2\theta-m_\pi^2(1-\cos^2\theta)]-16 m_f^2 m_\pi^2 \cos^2\theta}{s^2},
\label{eq:matrixffeepi}
\ee
where $m_{\pi}=139.6 \, \text{MeV}$~\cite{Beringer:1900zz} and  we include the form factor
\be
F_\pi(s)\approx\frac{1.20\ m_\rho ^2}{m_\rho^2-s-i m_\rho\Gamma_{\rho\pi\pi}},
\ee
from~\cite{Achasov:2011ra,Kroll:1967it}
where $m_\rho=775.26\, \text{MeV}$ is the mass of the $\rho(700)$ meson~\cite{Beringer:1900zz}, and $\Gamma_{\rho\pi\pi}\approx 149.1 \,\text{MeV}$ its decay width into pions~\cite{Beringer:1900zz}. This is the simplest form of $F_\pi(s)$ based on the vector dominance model~\cite{Achasov:2011ra,Kroll:1967it}.

The contribution of plasmon decay~\eqref{plasmon decay} to $\Gamma$, i.e. the energy density transferred per unit time to the DS, takes the form 
 \be
 \W_{\gamma^*\to \bar f f}=\sum _{\rm pol}\int \frac{d^3k}{(2\pi)^3}
 \left(\frac{1}{e^{\omega /\TG}-1}-\frac{1}{e^{\omega/\TDS}-1}\right)\omega\ \Gamma_{\gamma^*}
 \ee 
 where the plasmon decay rate in the comoving frame is 
 \be
 \label{plasmondecay}
\Gamma_{\gamma^*}=\frac{\alpha\, \epsilon^2}{3 \omega}Z
\left(m_\gamma^2+2m_f^2\right)\sqrt{1-\frac{4m_f^2}{m_\gamma^2}},
 \ee
 where $m_\gamma$ is the photon plasma mass defined by the dispersion relation $\omega^2-k^2=m_\gamma^2$ and $Z=Z(\omega,k)$ is the renormalization factor~\cite{Raffelt:1996wa}. 
We have to sum over photon polarizations, the two transverse and the longitudinal, for which $m_\gamma$ and $Z$ are different. We are interested in a plasma made of relativistic particles where transverse plasmons dominate the decay rate~\cite{Braaten:1993jw} so we neglect the longitudinal mode and use $\omega\gg m_\gamma$. 
In this case the plasma mass at first order in $\alpha$ is 
 \be
m_\gamma^2=\sum_i g_i Q_i^2 \frac{4\alpha}{\pi}\int_0^\infty\dd p  f_i(p) p
 \ee
where the sum goes over all charged particle species, $g_i$ controls the spin and color multiplicity of the particle and $Q_i$ is its electric charge. 
The renormalization factor is $Z\sim 1$ unless $\omega \sim m_\gamma$. 

\subsubsection{Decoupling channels}

While SM particle annihilation \eqref{eq:SMannihilation} and plasmon decay \eqref{plasmon decay} decrease approximately like $\text{exp}(-2 m_f/T)$ once $T\sim m_f$, Compton scattering \eqref{eq:comptscat} and Coulomb scattering \eqref{eq:coulomb} only decrease as $\text{exp}(-m_f/T)$. These channels are therefore prone to dominate when $T\lesssim m_f$. However, they are only important if $\TDS \sim \TG$ since they need a sizable abundance of MCPs to be effective. 

For Compton scattering~\eqref{eq:comptscat} we use the following matrix element in the rest frame of the MCP~\cite{Peskin}

\be
 |\widetilde{\mathcal{M}}|^2_{f\gamma \rightarrow f \gamma'}=8 (g'e\epsilon)^2 \left(\frac{\omega'}{\omega}+\frac{\omega}{\omega'}-\sin^2\theta\right),
\ee
where $\omega$ is the angular frequency of the incoming photon and $\omega'$ is the angular frequency of the outgoing HP. These two frequencies are related by Compton's formula
\be
\omega'=\frac{\omega}{1+\frac{\omega}{m_f}(1-\cos \theta)}.
\ee

The calculation of $\W$ for Coulomb scattering~\eqref{eq:coulomb} is quite cumbersome due to the forward Coulomb divergence. In vacuum we find~\cite{Hahn:2000kx}
\be
\label{eq:WCoulomb}
 |\widetilde{\mathcal{M}}|^2_{f e^- \rightarrow f e^-}=8 e^4\epsilon^2 \frac{(m^2 - s)^2 +  2 m^2 t + (m^2 - u)^2}{t^2},
\ee
where $s,u,t$ are the Mandelstamm variables and $m$ is the highest mass involved in the process. 
For pions we find
\be
\label{eq:WCoulombPi}
|\widetilde{\mathcal{M}}|^2_{f \pi^- \rightarrow f \pi^-}=8e^4\epsilon^2\frac{-s\, u + m_f^2 (s+u) +m_{\pi}^4-m_f^4}{t^2},
\ee
where $m_\pi$ is again the charged pion mass. We include a form factor \cite{Dominguez:2009cg}
\be
F_\pi(t)\approx\frac{m_\rho ^2}{m_\rho^2-t}.
\ee
Note that in the Coulomb energy transfer integral we neglect the mass of the lighter particle, which slightly underestimates $\W$ when both masses are similar. 

Matrices \eqref{eq:WCoulomb} and \eqref{eq:WCoulombPi} diverge at low $t$. 
The energy transfer, $E_\text{trans}$, has to vanish at $t=0$ so $E_\text{trans}\propto t$ at low $t$ and, after the $t$-integration, $\W$ is logarithmically sensitive to the cut-off. 
Since energy transfer via Coulomb scattering is most important when the MCP is decoupling and hence in the verge of being non-relativistic, we can take the cut-off to be the Debye screening momentum $k_\text{D}$ which in a relativistic plasma is just $m_\gamma$. 
We implement this minimum plasma screening by the substitution $t^2\to (t-m_\gamma^2)^2$ in \eqref{eq:WCoulomb}. 
We have checked the validity of that prescription in a few cases of interest by comparing with the energy transfer of a massive fermion in a QED plasma calculated in thermal-field theory including dynamical screening~\cite{lebellacTFT}. 
Since all SM particles can scatter on the MCP, we include all the channels discussed in~\ref{prod}.

\subsubsection{Other channels}

Another channel we include is the vector boson fusion process~\eqref{eq:annihi}. It is neither important for $\TDS \ll \TG$ nor for $\TDS < m_f$ but gives corrections for $\TG \sim m_f$. We include this channel using the following matrix element~\cite{Peskin}
\be
|\widetilde{\mathcal{M}}|^2_{\gamma \gamma' \rightarrow f\bar f}=8\epsilon^2e^4\frac{s^2(\cos^4{\theta}-1)+16 m_f^4(\cos^4{\theta}-2\cos^2{\theta}+2)-8m_f^2s(\cos^4{\theta}-\cos^2{\theta}+1)}{[s+(4m_f^2-s)\cos^2{\theta}]^2}.
\ee
 Processes \eqref{eq:annihi2} and \eqref{eq:comptscat2} are of order $\mathcal{O}( \epsilon^4)$ and can be neglected unless the MCPs are extremely light~\cite{Melchiorri:2007sq}, a case already excluded 
 by stellar evolution except in more involved models~\cite{Masso:2006gc} which we do not considered here. 

\subsection{Initial conditions}\label{initialconditions}

To compute the temperature ratios with the system of equations \eqref{system1}, we have to specify the initial conditions. In the spirit of a hidden sector we assume that the DS is absent after reheating ($\TDS=0$) and is dynamically created by SM reactions. As soon as some MCPs are produced via \eqref{eq:SMannihilation} and \eqref{plasmon decay}, they can generate HPs via $f \bar f \rightarrow \gamma' \gamma'$ until the distributions of both MCPs and HPs are thermal with a common temperature $\TDS$. 
Let us explore when this reaction is effective with a simple order-of-magnitude analysis, leaving $\mathcal{O}(1)$ factors aside.   
At high $\TG$, SM fermion pair annihilation creates a population of MCPs of density $n_\text{MCP}\sim \alpha^2\epsilon^2 \TG^4/H$ and typical momentum $\mathcal{O}(\TG)$. The reaction $f \bar f \rightarrow \gamma' \gamma'$ starts to be effective when this population is enough to ensure that one MCP suffers one annihilation per Hubble time, i.e. when 
\be
\sigma (f \bar f \rightarrow \gamma' \gamma')n_\text{MCP}\frac{1}{H}\approx
\frac{g'^4}{16\pi^2 \TG^2} \frac{\alpha^2\epsilon^2 \TG^4}{H}\frac{1}{H}\approx
\frac{g'^4 \alpha^2\epsilon^2M_p^2}{16\pi^2 \TG^2} 
\sim 1, 
\ee
which happens at 
\be
\label{DSthermal}
\TG\sim \frac{g'^2 \alpha \epsilon M_p }{4\pi} =10^7\ {\rm GeV}\ \left(\frac{g'}{0.1}\right)^2
\frac{\epsilon}{10^{-8}} . 
\ee
Once this temperature is reached, the typical momentum of the DS particles is degraded 
due to the thermalization to $\TDS<\TG$ and the cross section $\sigma (f \bar f \rightarrow \gamma' \gamma')$ 
increases, boosting the process.  
Thus $g'$ is sizable, this thermalization happens so fast that assuming a common temperature $\TDS$ is justified.

From a numerical point of view, $\TDS=0$ is a difficult initial condition since the DS is populated extremely fast in the first Hubble times. 
This urge stops when we reach the regime 
\be
4H\rho (\TDS) = \W (\TG,\TDS, \epsilon,m_f). \label{eq:initial}
\ee
One can show that the higher $\TG$  the less our results will depend on the initial conditions. We, therefore, start our calculations at $\TG \sim 10^7 \ \text{GeV}$ and compute the initial condition $\TDS$ for given values of $\epsilon$ and $m_f$ using eq.~\eqref{eq:initial}.

For MCPs with $\epsilon\gtrsim 10^{-4}$ in the mass range of interest the DS and SM thermalize so fast  that setting $\TDS=\TG$ is equivalent to starting with $\TDS=0$. For such large kinetic mixings we will therefore assume thermalization of the SM sector with the DS.

\subsection{Numerical evaluation}

Now that the initial conditions and the source term are fixed, we can track the temperature ratios down to the CMB epoch for different MCP masses and minicharge $\epsilon$ using \eqref{system1}. 
In the strong coupled regime ($\epsilon> 10^{-4}$) we linearize the source term $\W$ around $T_\gamma=\TDS$ to improve the stability of the solver. We can then calculate \neff with eq.~\eqref{CMB formula}. The error in $\Delta \neff$ introduced by the linearization appears to be $\sim 5 \%$ for the phenomenologically interesting region of $m_f> 100\ \text{MeV}$ (see section \ref{BBN}).
As a cross check of our simulations, we have compared to~\cite{Davidson:2000hf}. We reproduce their results when we reduce our reaction set to plasmon decay and pair annihilation only. 

\section{Results for \neff at the CMB epoch}

The MCP model we consider here has three parameters: $m_f$, $\chi$ and $g'$. It is more convenient to use $\epsilon=g'\chi/e$ instead of $\chi$ because this is the  parameter that controls the energy transfer between the SM and DS in the early thermalization of the DS and during decoupling. 
Of the $m_f,\epsilon,g'$ set, the hidden coupling $g'$ is perhaps the least relevant. It controls the thermalization of the DS by itself ---but the requirements are not very restrictive--- and affects the decoupling through Compton scattering, which is dominant only for $m_f\lesssim m_e$ unless $g'$ is large.   
Note that, of all the reactions listed in Eqs.~\eqref{eq:SMannihilation}-\eqref{eq:coulomb} only Compton~\eqref{eq:comptscat} and vector fusion~\eqref{eq:annihi} depend explicitly on $g'$. 
Thus we have decided to scan the parameter space $m_f,\epsilon$ for just three representative values of $g'=1,0.1,0.01$. 
Based on these cases, we can extend our conclusions to further values of $g'$.

\begin{figure}[t] \centering
 \includegraphics[width=0.65\textwidth]{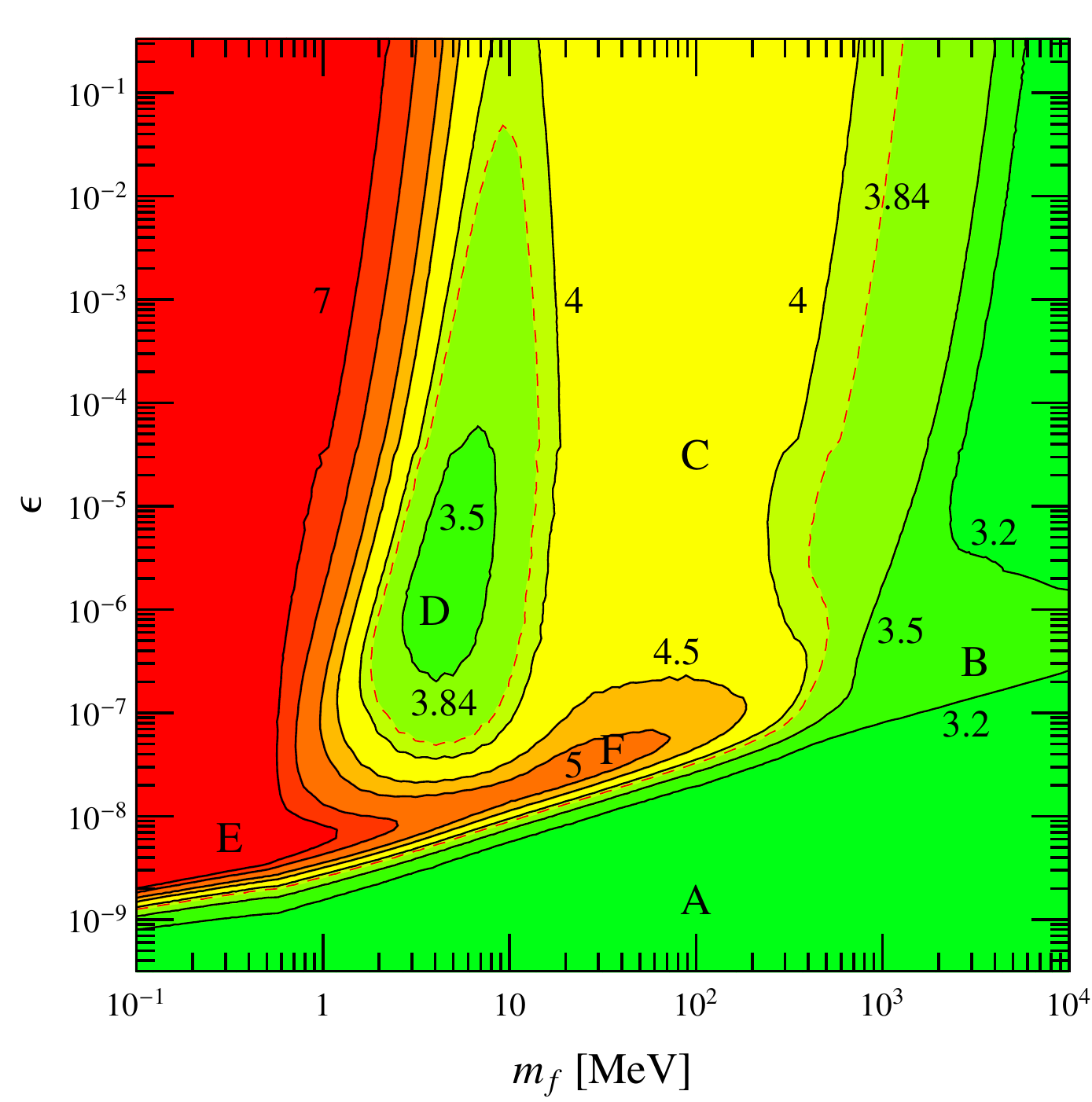}
 \caption[Result for $\neff$ at CMB for $g'=0.1$]
 {{\small{Isocontours of $\neff$ at the CMB epoch as a function of the hidden fermion mass $m_f$ and the minicharge $\epsilon$ for a value of the hidden gauge charge $g'=0.1$. 
Regions denoted with letters are further discussed in the main text. 
Dark green coloring denotes regions close to the SM value $\neff=3$. 
Light green and yellow regions lie between $3.5-4.5$. 
Orange and red denote higher values $\neff>4.5$. 
The red dotted line shows the 95\% upper exclusion limit $\neff=3.84$ (Planck+WP+highL+BAO) by Planck~\cite{Ade:2013zuv}.}}} \label{fig:Neffg-1} \end{figure}

\subsection{Exemplary case: $g'=0.1$}

Our results for $g'=0.1$ are shown in figure~\ref{fig:Neffg-1}, which displays the isocontours of \neff in the $m_f-\epsilon$ plane. The general structure follows from simple considerations. 
At large $\epsilon$ MCPs and HPs thermalize very soon with the SM and their final contribution to $\neff$ depends mostly on the moment of decoupling, set by the MCP mass, but not much on the strength of the coupling. Thus for large $\epsilon$,  isocontours are vertical. 
For sufficiently low $\epsilon$, the DS does not reach a significant abundance and $\neff\to 3$. 
The boundary is given by those MCPs which had almost thermal abundance when $\TG\sim m_f$ and the DS thermalization process is quenched. 
The ratio of the DS and SM bath energy densities is $\propto \epsilon^2/\TG$ at early times (before an eventual thermalization)\footnote{The rates of MCP production processes are proportional to $\epsilon^2$. The $1/\TG$ factor is derived in section~\ref{prod}.} and decouples as $\propto \epsilon^2/m_f$. Saving $\mathcal{O}(1)$ factors, 
$\neff$ is different from 3 only if the DS abundance is comparable to the SM, which gives us the requirement $\epsilon^2/m_f \gtrsim \rm const.$ and the rough slope of the lowest isocontours $\epsilon\propto \sqrt{m_f}$. 

To elaborate on the physics responsible for the patters in figure~\ref{fig:Neffg-1}, we discuss in the following six different peculiar regions labelled with capital letters. For each of them we show the evolution of the neutrino and DS temperatures in Figs.~\eqref{fig:CMBD}-\eqref{fig:CMBE}. 


\begin{figure}[p] \begin{center}
$\begin{array}{cc}
    \includegraphics[width=0.47\textwidth]{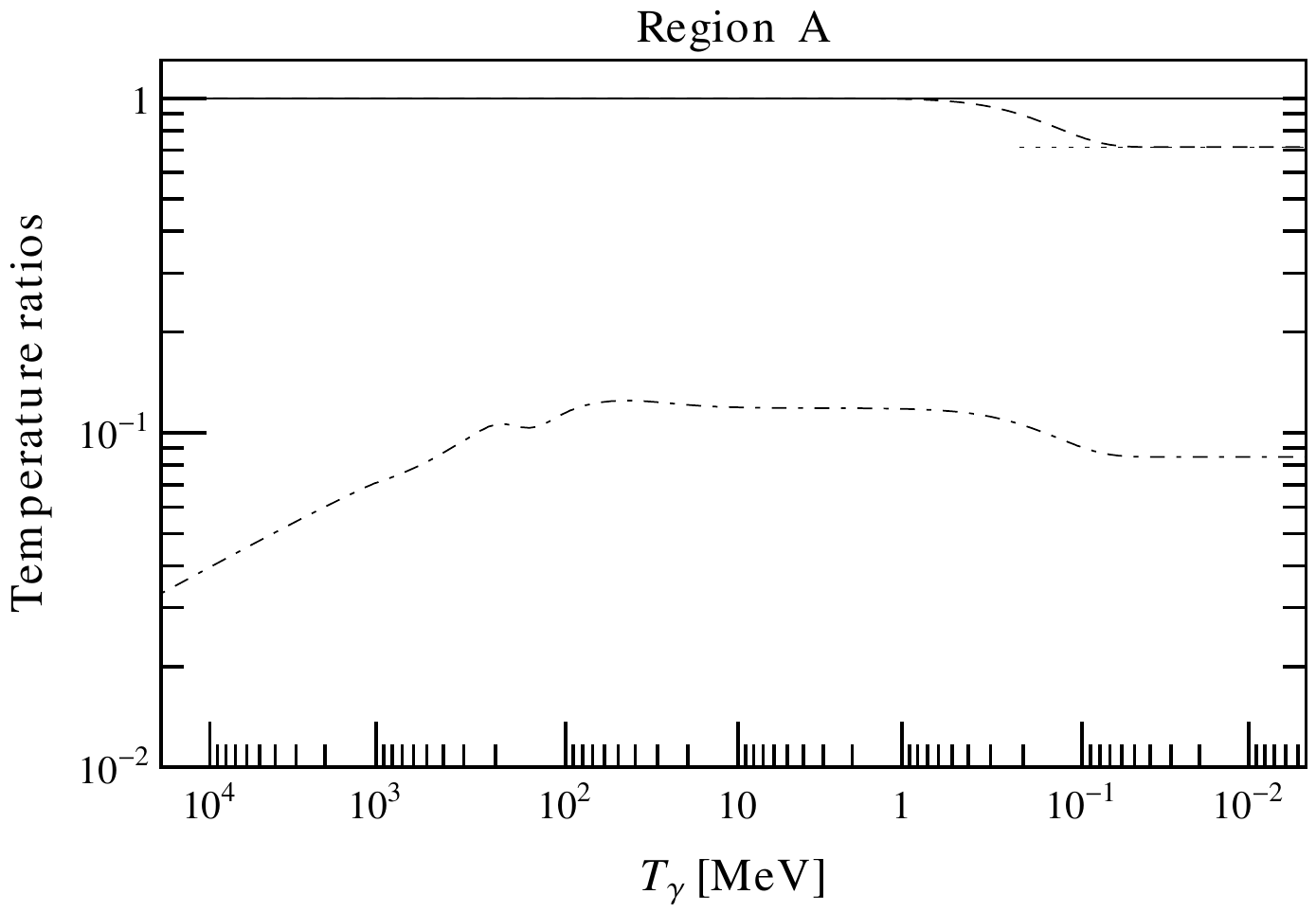}     &
    \includegraphics[width=0.47\textwidth]{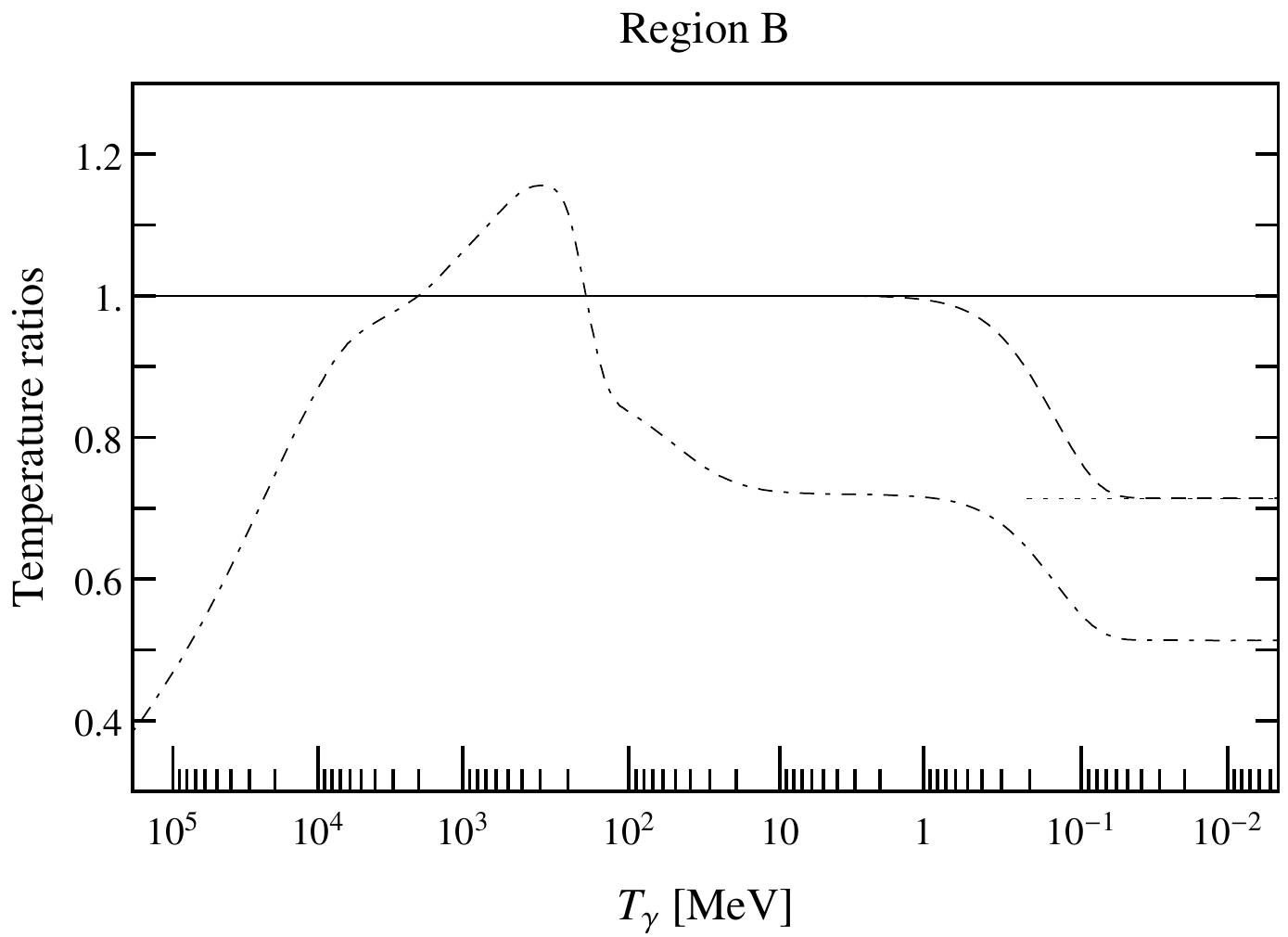}
 \end{array}$\end{center} 
 \vspace{-.61cm}
 \caption[Region A \& B for CMB]{
 {\small{Evolution of the temperature ratios $T_\nu/\TG$ (dashed) and $\TDS/\TG$ (dot-dashed). The black line at 1 denotes thermalization with the SM. \emph{Left:} region A ($m_f=0.1 \ \text{GeV}$, $\epsilon=10^{-9}$). \emph{Right:}  region B ($m_f=3.15 \ \text{GeV}$, $\epsilon=3 \times 10^{-7}$).}}} \label{fig:CMBA} \end{figure}

   \begin{figure}[p] \begin{center} 
$\begin{array}{cc}
    \includegraphics[width=0.47\textwidth]{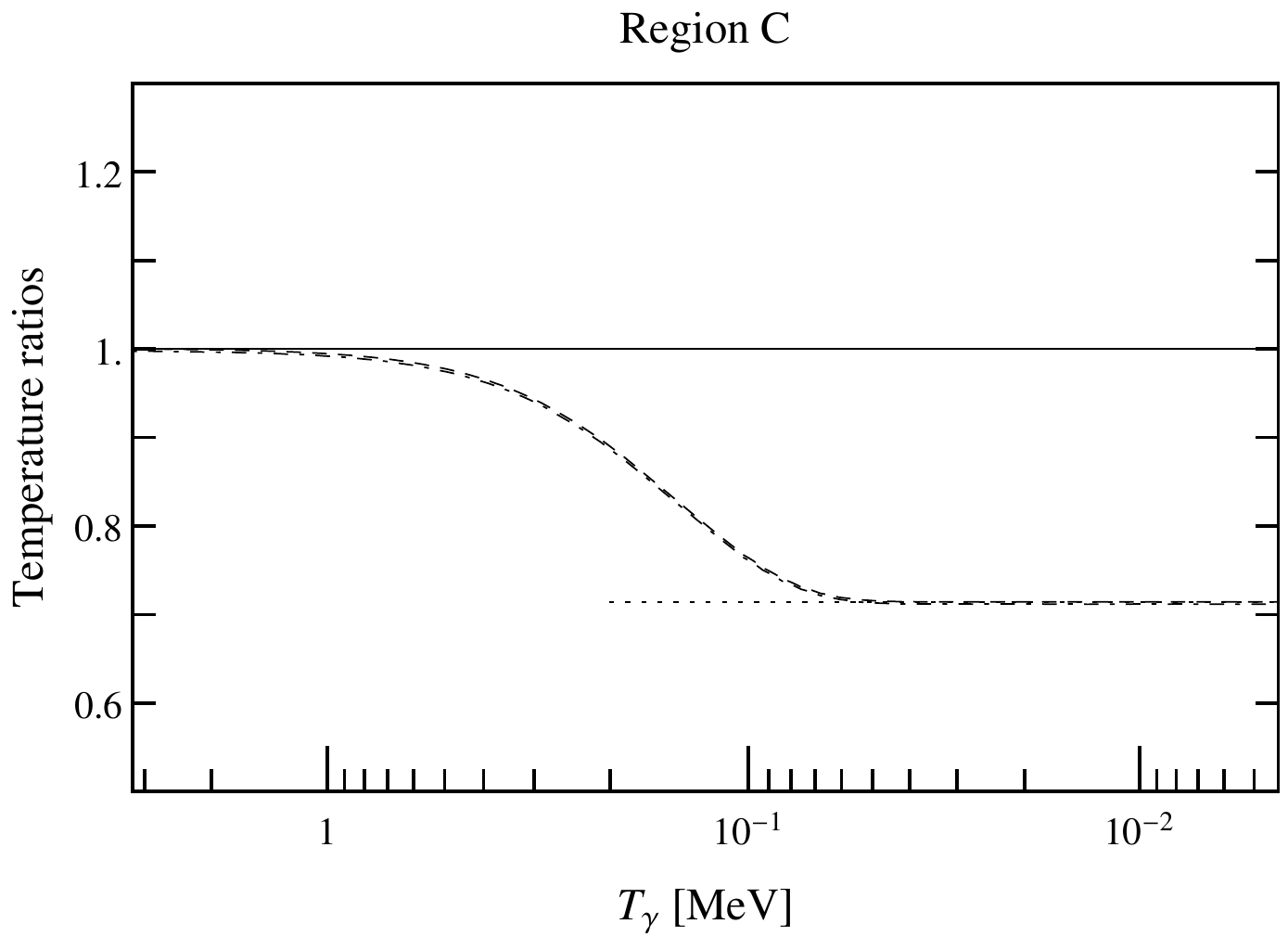} &
    \includegraphics[width=0.47\textwidth]{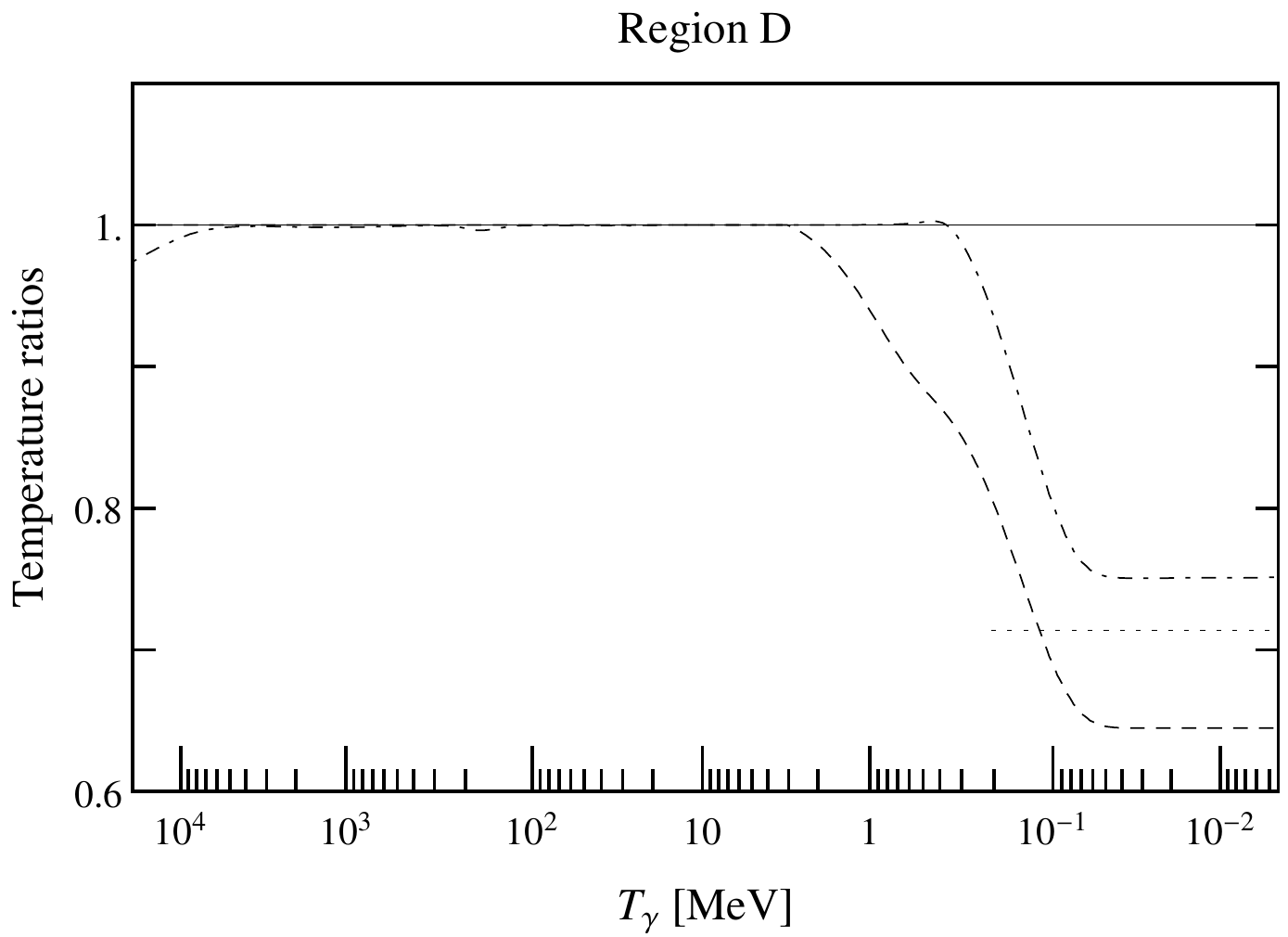}
 \end{array}$\end{center}
  \vspace{-.61cm}\caption[Region C \& D for CMB]{{\small{Same as in figure\eqref{fig:CMBA}. \emph{Left:} region C ($m_f=100 \ \text{MeV}$, $\epsilon=3 \times 10^{-5}$). The dashed and dash-dotted lines lie on top of each other. \emph{Right:}  region D ($m_f=3 \ \text{MeV}$, $\epsilon=10^{-6}$).}}}
 \label{fig:CMBD} \end{figure}

\begin{figure}[p] \begin{center}$
\begin{array}{cc}
    \includegraphics[width=0.47\textwidth]{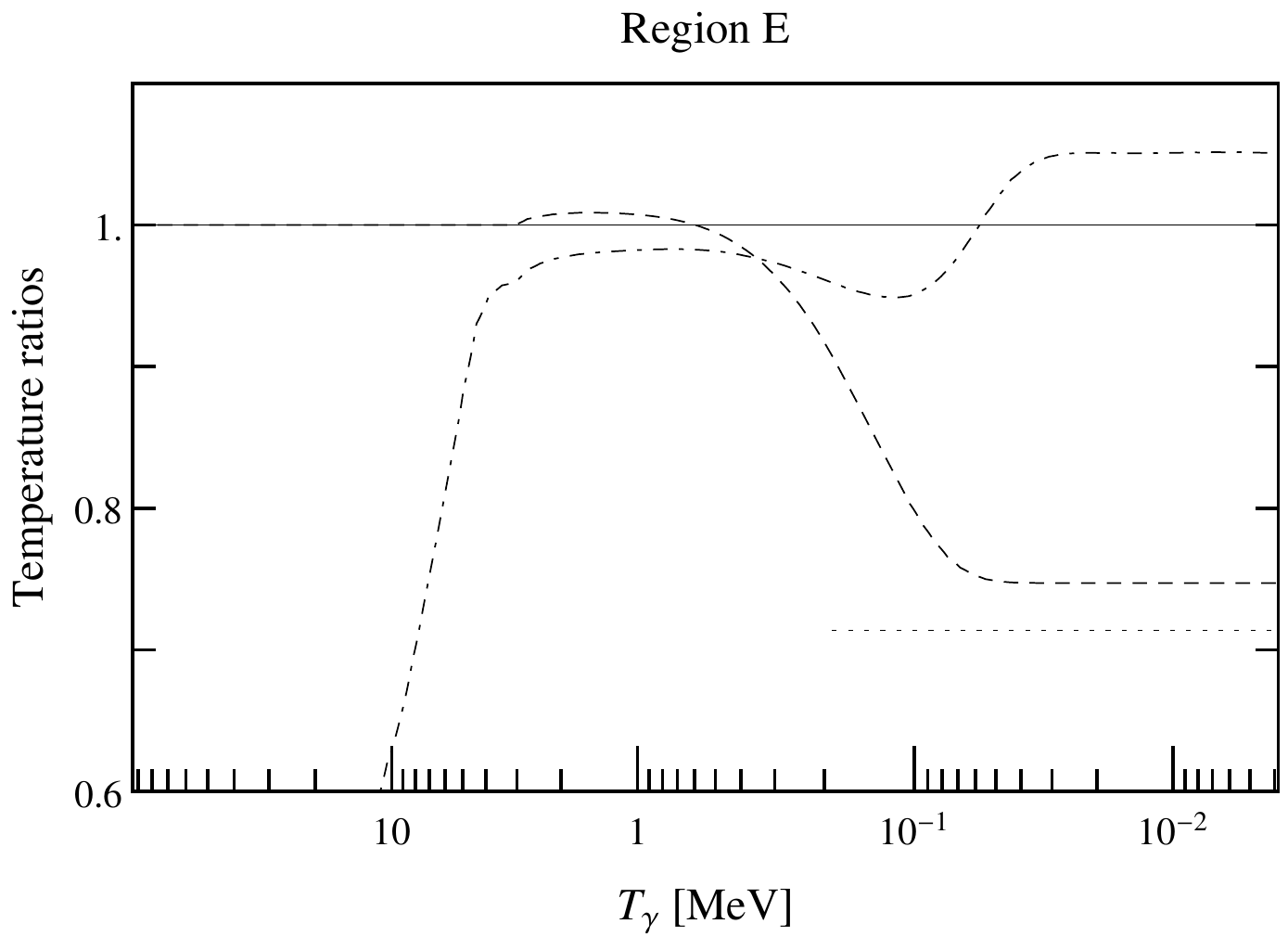} &
    \includegraphics[width=0.47\textwidth]{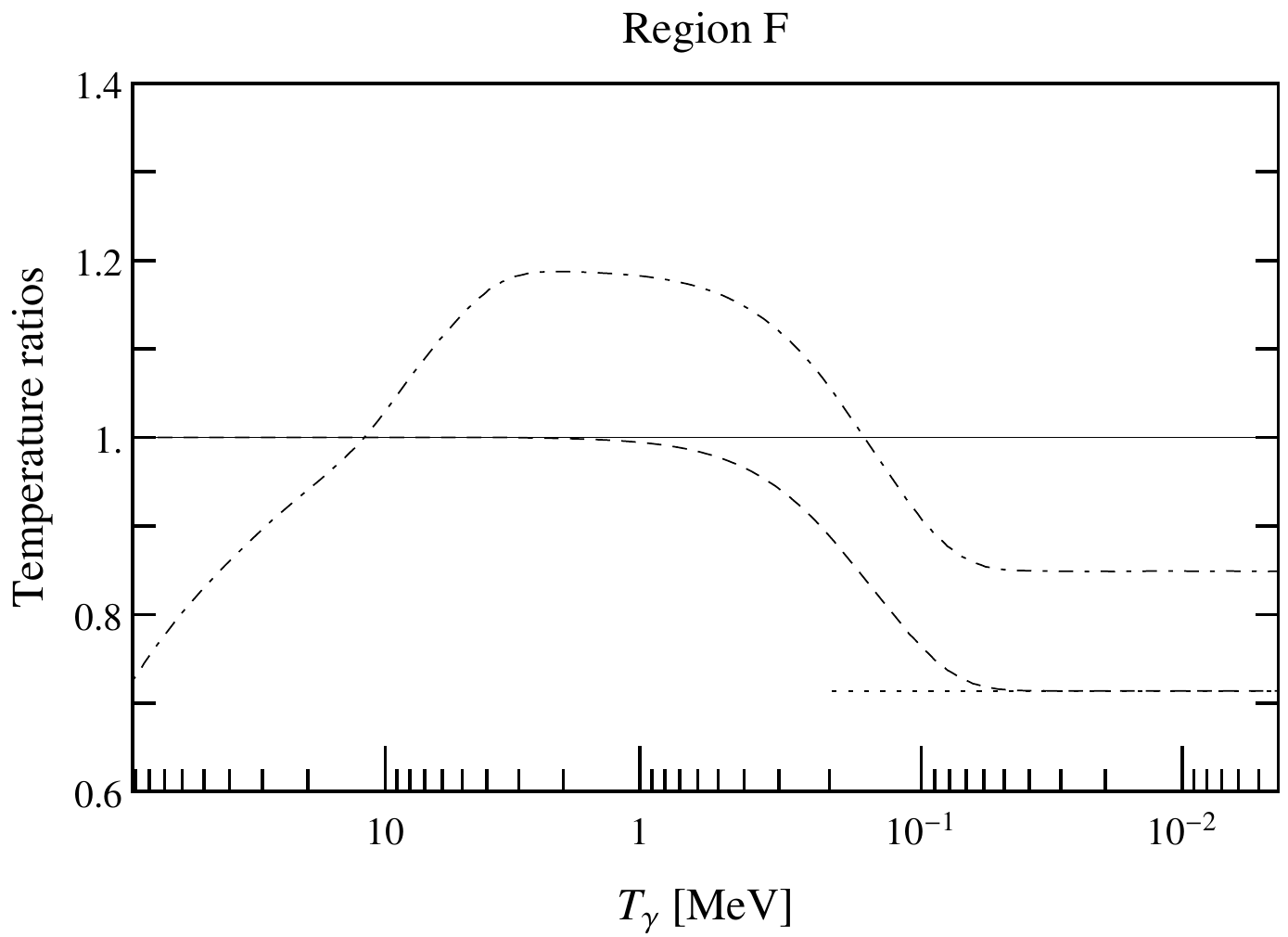}
 \end{array}$
\end{center} 
\vspace{-0.6cm} \caption[Region E \& F for CMB]{{\small{Same as in figure\eqref{fig:CMBA}. \emph{Left:} region E ($m_f=0.3 \ \text{MeV}$, $\epsilon=6\times 10^{-9}$). \emph{Right:} region F ($m_f=31 \ \text{MeV}$, $\epsilon=3 \times 10^{-8}$).}}} \label{fig:CMBE}\end{figure}
\afterpage{\clearpage}
\begin{itemize}
 \item Region A (figure~\ref{fig:CMBA} left)
  
In this region, the minicharge is so small that the DS never reaches equilibrium with the SM and thus \neff does not deviate from the standard value.
We see that $\TDS/T_\gamma$ (dot-dashed line) increases slowly until $T_\gamma \sim m_f$ (in this particular case $m_f=100$ MeV) when the pair creation of MCPs becomes exponentially suppressed due to the lack of energetic-enough SM charged particles. At this moment $\TDS$ is only $\sim T_\gamma/10\sim m_f$, the thermal population of MCPs becomes soon exponentially suppressed and the DS completely decoupled (no Compton or Coulomb processes are efficient because the MCP population is tiny).  At $T_\gamma\sim 3$ MeV neutrinos decouple and the subsequent $e^\pm$ annihilation heats the photon bath with respect to the decoupled neutrinos and HPs . This makes the temperature ratios $T_\nu/T_\gamma$ (dashed line) and $\TDS/T_\gamma$ drop between $T_\gamma\sim $1 and 0.1 MeV.  
The final temperature of neutrinos is the standard value $T_\nu/T_\gamma=(4/11)^{1/3}$. We marked this value as a dotted horizontal line here and in all the plots.  
 \item Region B (figure~\ref{fig:CMBA} right); 
 
For substantially larger couplings, the DS reaches equilibrium with the SM sector. 
If $m_f> 1 \, \text{GeV}$, the DS freezes out when the light quarks and gluons are still present in the SM bath.   
These degrees of freedom will eventually heat up the SM bath with respect to the HP.  
The temperature of neutrinos has its standard history. 
In the particular case depicted, the DS and SM sectors reach thermal equilibrium when $\TG\sim m_f$, i.e. 
very close to their decoupling. The coupling of the sectors is still weak so that most of the MCP entropy heats the
HP bath with respect to the SM. Thus, for a little while, until the QCD phase-transition triggers the disappearance of colored degrees of freedom, the HPs have a higher temperature than the SM. 
%
 \item Region C (figure~\ref{fig:CMBD} left)
 
For large $\epsilon$ and $m_f$ in the approximate range $(10-1000)$ MeV, the DS decouples after the QCD phase transition but before neutrino decoupling. 
In this temperature range the only SM particles that can heat the SM bath with respect to DS after decoupling are  electrons, muons and pions, whose number of degrees of freedom are comparable to the DS. Thus, $\TDS$ ends up being close enough to $\TG$ to contribute sizably to $\neff$. 
Below $m_f\sim 100$ MeV only electrons are relevant. The case depicted in figure~\ref{fig:CMBD} shows such a case. 
The MCPs have decoupled at $\TG\sim m_f =100$ MeV in thermal equilibrium with the SM so the SM and HPs have the same temperature. Later, the $e^\pm$ annihilation epoch heats photons, but not neutrinos nor HPs which share the same temperature $(4/11)^{1/3}\TG$. 
In this range we thus have typically $\neff\sim 3+\frac{8}{7}= 4.14$. 

 \item Region D (figure~\ref{fig:CMBD} right)
 
For smaller $m_f\in (1-10)$ MeV, the DS decouples before $e^\pm$ annihilation but after neutrino decoupling. 
When the MCPs become non-relativistic, they deposit a part of their energy in the SM plasma. 
Thus, $\TG/T_\nu$ becomes larger than in the standard scenario, and consequently the neutrinos contribute less than 3 to \neff. 
In figure~\ref{fig:CMBD} this is evident from the fact that $T_\nu/\TG$ becomes smaller than $(4/11)^{1/3}$. 
HPs get some of the MCP energy and might even get their share also from $e^\pm$'s so they have a sizable contribution to \neff. 

\item Region E (figure~\ref{fig:CMBE} left)
 
For $m_f<m_e$ and $\epsilon>2\times 10^{-9}$, very high values for $\neff$ are realized. 
The reason is that $e^\pm$ annihilation pumps energy not only into photons but into the DS as well.   
If the DS is weakly coupled, most of the MCP energy goes into HPs at decoupling and therefore we have $T_\nu/\TG>(4/11)^{1/3}$ and a large HP contribution. An example of this is depicted in figure~\ref{fig:CMBE}. 
If the coupling is strong, the MCP annihilation dumps exactly the amount of energy needed to restore the standard value for $T_\nu/\TG$ into the photon bath (a consequence of dealing with a fermionic Dirac MCP, which has the same degrees of freedom as electrons and positrons). 

 \item Region F (figure \ref{fig:CMBE} right)
 
In this region, the SM-DS interactions are strong enough to bring the DS energy to values close to thermalization, but not sufficient to ensure the fast transfer of energy between the sectors when the MCP becomes non-relativistic and decouples. Thus most of the energy of the MCPs ends up in HPs, which get hotter than photons at least for some period of time. In figure~\ref{fig:CMBE} we depict such a case. MCP decoupling happens before neutrino decoupling so that the neutrino density adopts its standard value. 
$\TDS/\TG$ is higher than 1 after decoupling, but is eventually suppressed by $e^\pm$ annihilation to end up falling below 1. Nevertheless, the DS contribution to $\neff$ can be still quite sizable. In this example only the HP contributes as 2 effective neutrinos. 
 
\end{itemize}

\subsection{Cases $g'=1$ and $g'=0.01$} 
 
Most of the reaction rates depend on the combination $ (e \epsilon)^2\equiv (g'\chi)^2 $ so a change in $g'$ can always be compensated by varying $\chi$ accordingly. This is the case for $e^\pm$ annihilation~\eqref{eq:SMannihilation}, plasmon decay~\eqref{plasmon decay} and Coulomb scattering~\eqref{eq:coulomb}, which are responsible for production and decoupling in most of the parameter space. 
On the other hand, Compton scattering~\eqref{eq:comptscat} and vector fusion~\eqref{eq:annihi} are proportional to $g'^2 (e\epsilon)^2$. Thus, increasing (decreasing) $g'$ increases (decreases) Compton scattering and MCP annihilation relative to Coulomb scattering, plasmon decay and $e^\pm$ annihilation for fixed $(m_f,\epsilon)$. 

Our results for $\neff$ can be seen in figure~\ref{fig:Neffg1} for $g'=1$ (left) and for $g'=0.01$ (right). 
The case $g'=0.01$ is virtually indistinguishable from the $g'=0.1$ case. This finding corroborates the fact that 
the Compton and MCP annihilation processes do not play a significant role in the value of \neff, at least for $g'\leq 0.1$. 
The only difference is a slight increase of \neff at low $\epsilon$ at the lowest masses $m_f\lesssim m_e$. Low mass  MCPs with almost thermal abundance can still mediate some energy into the DS by the Compton process after electrons have annihilated (when the Coulomb process is inefficient). 

The $g'=1$ case is more interesting. All the isocontours of \neff move down in $\epsilon$ by a factor $2\sim4$, depending on the region. This indicates that the Compton and MCP annihilation processes have become the dominant energy-transfer reactions in the DS-SM decoupling. 

\subsection{Implications from Planck}

\begin{figure}[t] \centering
\includegraphics[width=0.49\textwidth]{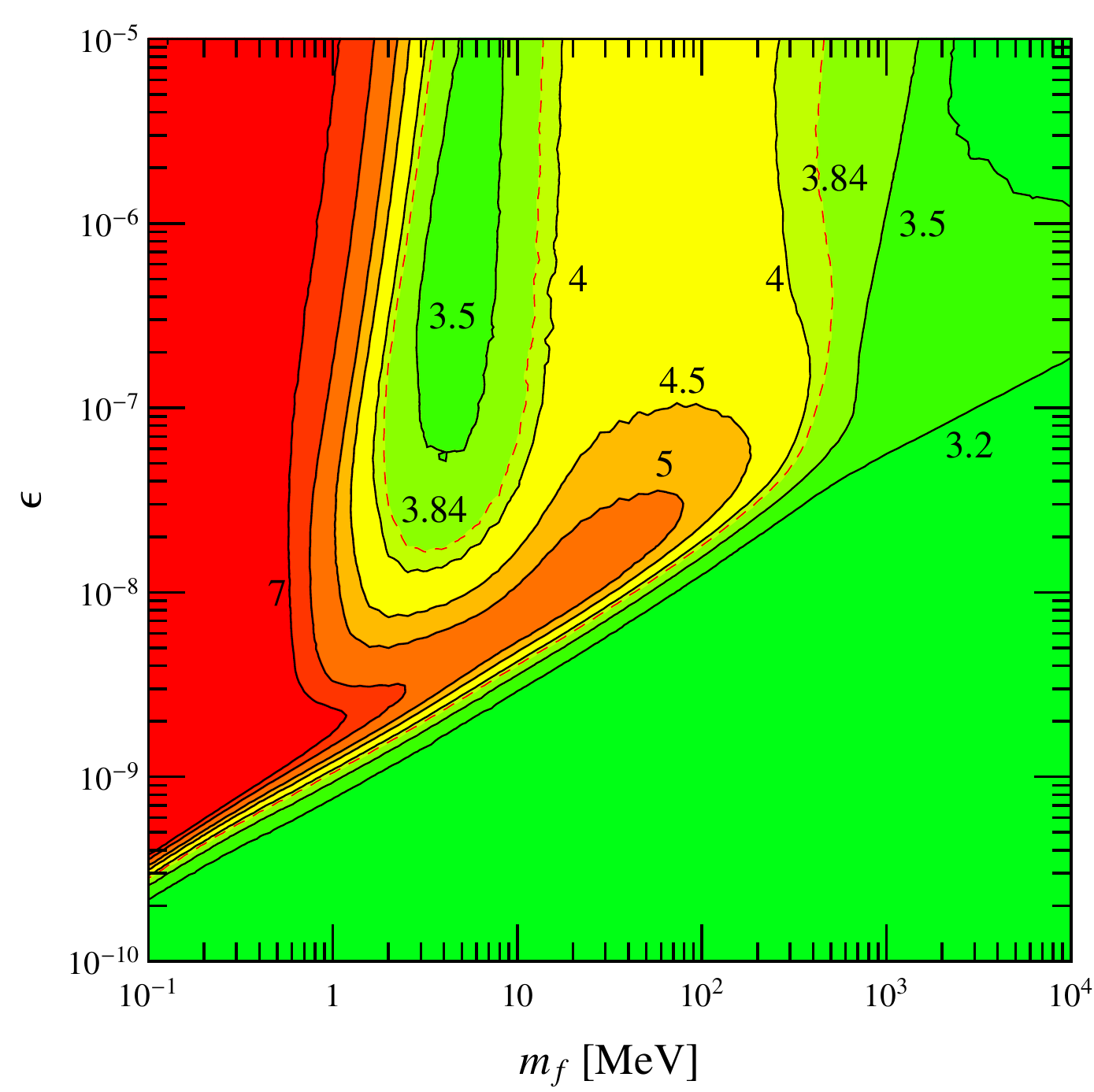}\hspace{-0.1cm}
\includegraphics[width=0.49\textwidth]{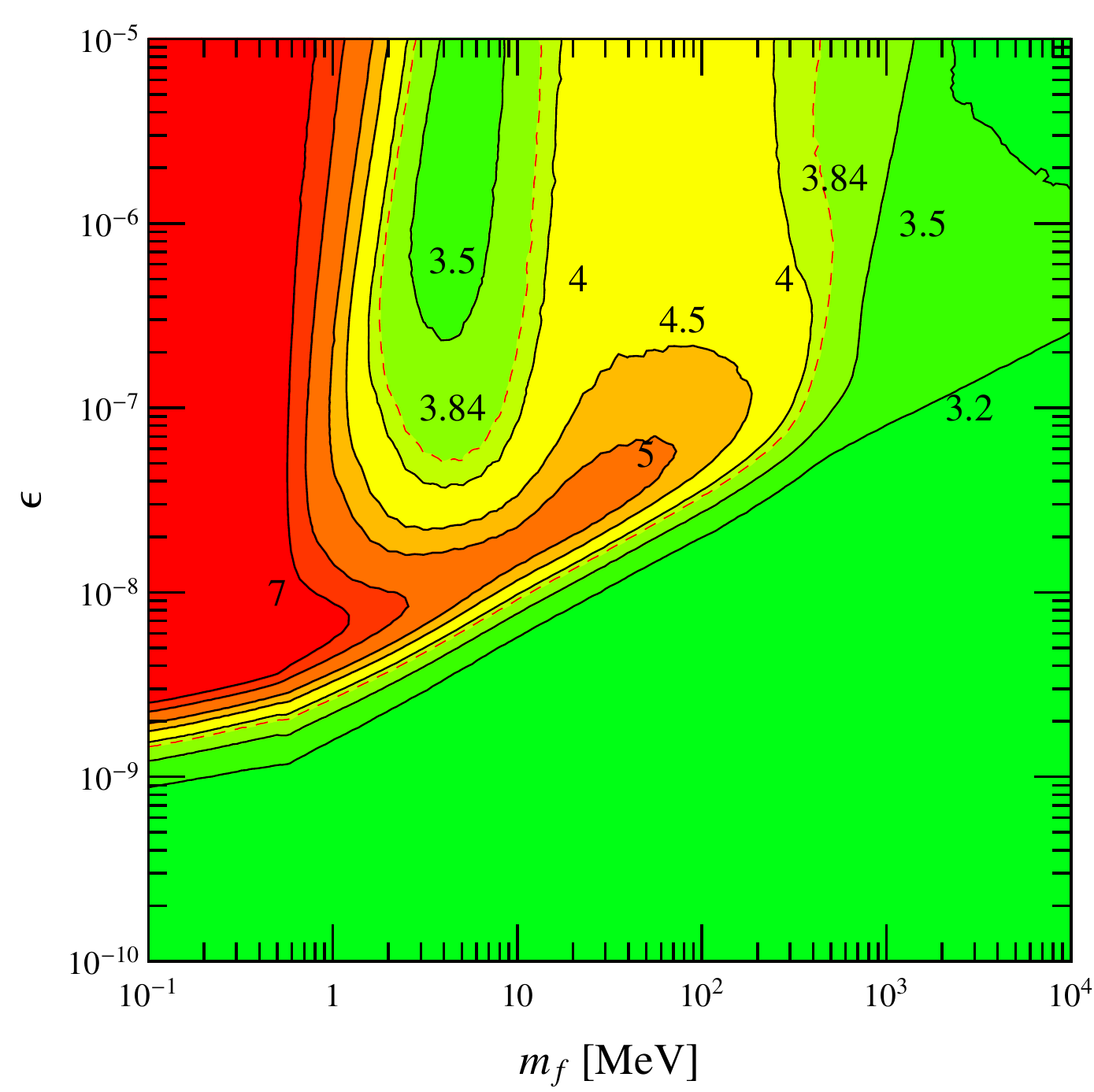}
\caption[Result for $\neff$ at CMB for $g'=1$]{{\small{
Isocontours of $\neff$ at the CMB epoch as a function of the hidden fermion mass $m_f$ and the minicharge $\epsilon$ for $g'=1$ (left) and $g'=0.01$ (right). The color coding is the same as in figure~\ref{fig:Neffg-1}. }}}
\label{fig:Neffg1} \end{figure}

The 95\% C.L. Planck upper limit, $\neff<3.84$ (Planck+WP+highL+BAO)~\cite{Ade:2013zuv}, is marked with a red-dashed line  in figures~\ref{fig:Neffg-1} ($g'=0.1$) and~\ref{fig:Neffg1} ($g'=1$ and $g'=0.01$). 
The constraints are independent of $g'$ for $g'\lesssim 0.1$ and thus  figure~\ref{fig:Neffg-1} is valid for $g'<0.1$ as well. 
Note that for low $g'$ one needs to check that the MCPs and HPs thermalize (the thermalization $\TG$ of eq.~\eqref{DSthermal} has to be larger than $m_f$ for our constraints to be consistent). 

Let us also recall that the MCP relic abundance behaves as self-interacting dark matter, which is excluded by a number of arguments to be a dominant component of the observed cold dark matter~\cite{McDermott:2010pa,Boehm:2001hm}.  For very small $g'$ the relic abundance can be significant and these bounds have to be taken into account. 

All in all, the Planck analysis disfavors MCPs with masses between $14 \ \text{MeV} < m_f < 390 \ \text{MeV}$ for a wide range of minicharges $\epsilon > 10^{-7}$. For larger minicharges, the bound improves so that $m_f < 1190 \ \text{MeV}$ is excluded for, e.g., $\epsilon = 10^{-1}$. Interestingly, a broad range of $\epsilon$ is favored in the $\sim 5$ MeV mass range. In the next section, we show that this region is however ruled out by BBN. 

We highlight the Planck result excluding the HST bias because it shows the potential of future \neff measurements to exclude robustly MCP masses up to GeV. The same spirit showed for instance in~\cite{Brust:2013ova}. Including HST data implies $\neff<4$ (Planck+WP+highL +BAO+HST) at 95\% C.L.~\cite{Ade:2013zuv}, which changes our results very little. 
We should however remark the extreme sensitivity of the MCP mass bound to $\neff$: relaxing the constraint to $\neff<4.2$ would shift the constraint to $m_f>1$ MeV or so as emphasized previously~\cite{Davidson:1993sj}. 
A novel result of this paper is that this is only true above $\sim 10^{-7}$. The large-\neff peninsula around $\sim 10^{-8}-10^{-7}$ would still be excluded. 
Let us once more remark that the discrepancy on the value of $H_0$ favored by CMB and the one implied by local measurements (HST) prevents to make strong claims about exclusion limits on \neff.

 \section{Constraints from big bang nucleosynthesis \label{BBN}}
 
The energy content of the Universe drives its expansion influencing the effectiveness of the nuclear reactions of big bang nucleosynthesis. The extra radiation predicted in the MCP+HP model described here increases the  expansion with respect to the standard case. In such a Universe electroweak reactions freeze-out earlier, which implies more neutrons during BBN, and BBN itself happens earlier in time (so less neutrons decay). Since eventually all neutrons end up forming part of $^4$He nuclei, our scenario implies a larger-than-standard $^4$He yield. 
 
 \begin{figure}[t] \centering
 \includegraphics[width=0.5\textwidth]{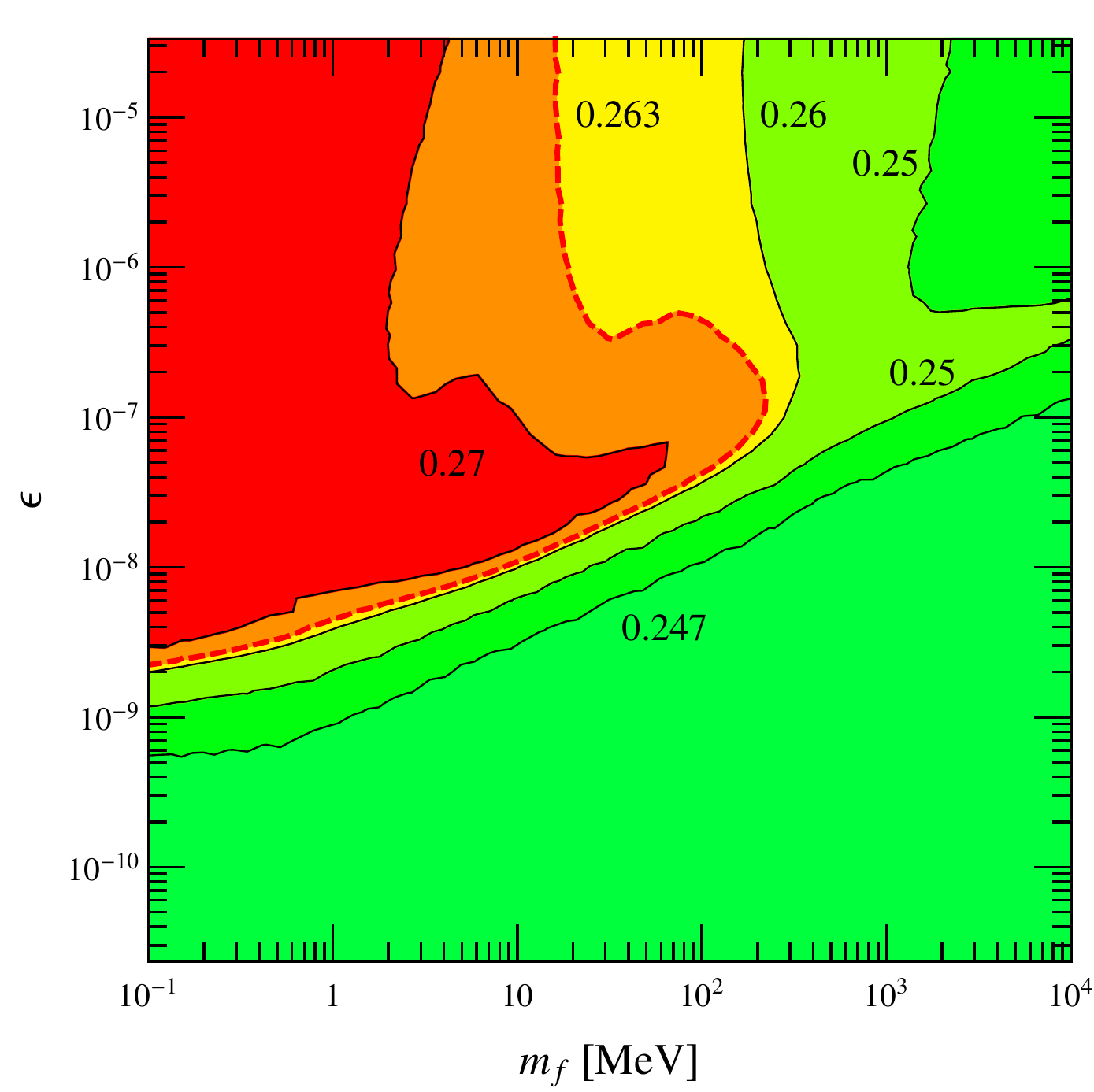}
 \caption[Helium yield $Y_p$]{\emph{\small{Helium yield $Y_p$ as a function of hidden fermion mass $m_f$ and minicharge $\epsilon$ for $g'=0.1$. Numbers correspond to the value of $Y_p$ of the closest contour line. Dark green coloring denotes regions far away from the upper limit $Y_p<0.263$~\cite{Mangano:2011ar}. The limit is given by the red dashed contour line. Orange and red regions are excluded on more than a $2\sigma$ level.}}} \label{fig:heliumg-1} \end{figure}

Note that MCPs and HPs do not affect \emph{directly} the relevant electroweak or nuclear reactions relevant in BBN, 
they do it only \emph{indirectly} by affecting \neff (and thus $H$) and the baryon to photon ratio $\eta$ (we have seen that MCPs and HPs can take and give entropy to the photon bath). There exist very accurate calculations of the relic abundance of primordial elements (helium, deuterium, lithium, ..) as a function of \neff and $\eta$, which are the only unknowns in the standard BBN scenario. However, when we include MCPs and HPs both \neff and $\eta$  \emph{can evolve} during the temperature ranges relevant for BBN ($T\sim $ 100 keV-2 MeV) and a simple rescaling of standard results is not always possible. 
Thus, we have adapted the BBN code employed in~\cite{Cadamuro:2011fd} to compute the primordial abundances of nuclei. As input we have the thermal histories that we computed in the previous section to evaluate \neff at the epoch of the CMB. 
We find that the $^4$He abundance gives an additional interesting constraint on the parameter space of the MCPs.  Isocoutours of the yield $Y_p=4 n_{\rm He}/n_B$ (normalized to the total baryon density) are shown in figure~\ref{fig:heliumg-1}. 
Using the constraint $Y_p<0.263$~\cite{Mangano:2011ar},  we can exclude MCPs with $m_f<16 \ \text{MeV}$ for $\epsilon > 1.4\times 10^{-8}$. This eliminates the region $2 \ \text{MeV} < m_f < 14 \ \text{MeV}$ still allowed by \neff at the CMB epoch. Note that this constraint is slightly more conservative than the recently suggested $Y_p=0.254\pm0.003$~\cite{Izotov:2013waa} (actually it corresponds to a 99\% C.L. exclusion).  
Hence the combination of BBN and Planck data disfavors MCPs with $\epsilon's$ in the range $10^{-7}-10^{-8}$  for masses $m_f < 390 \ \text{MeV}$.  

We checked that deuterium does not give us any further constraint. We do not consider lithium in this study since already in standard BBN the amount of $^7$Li differs from the SM prediction by more than $4\sigma$~\cite{Cyburt:2008kw, Fields:2011zzb}.
 
\begin{figure}[t] \centering
\includegraphics[width=0.49\textwidth]{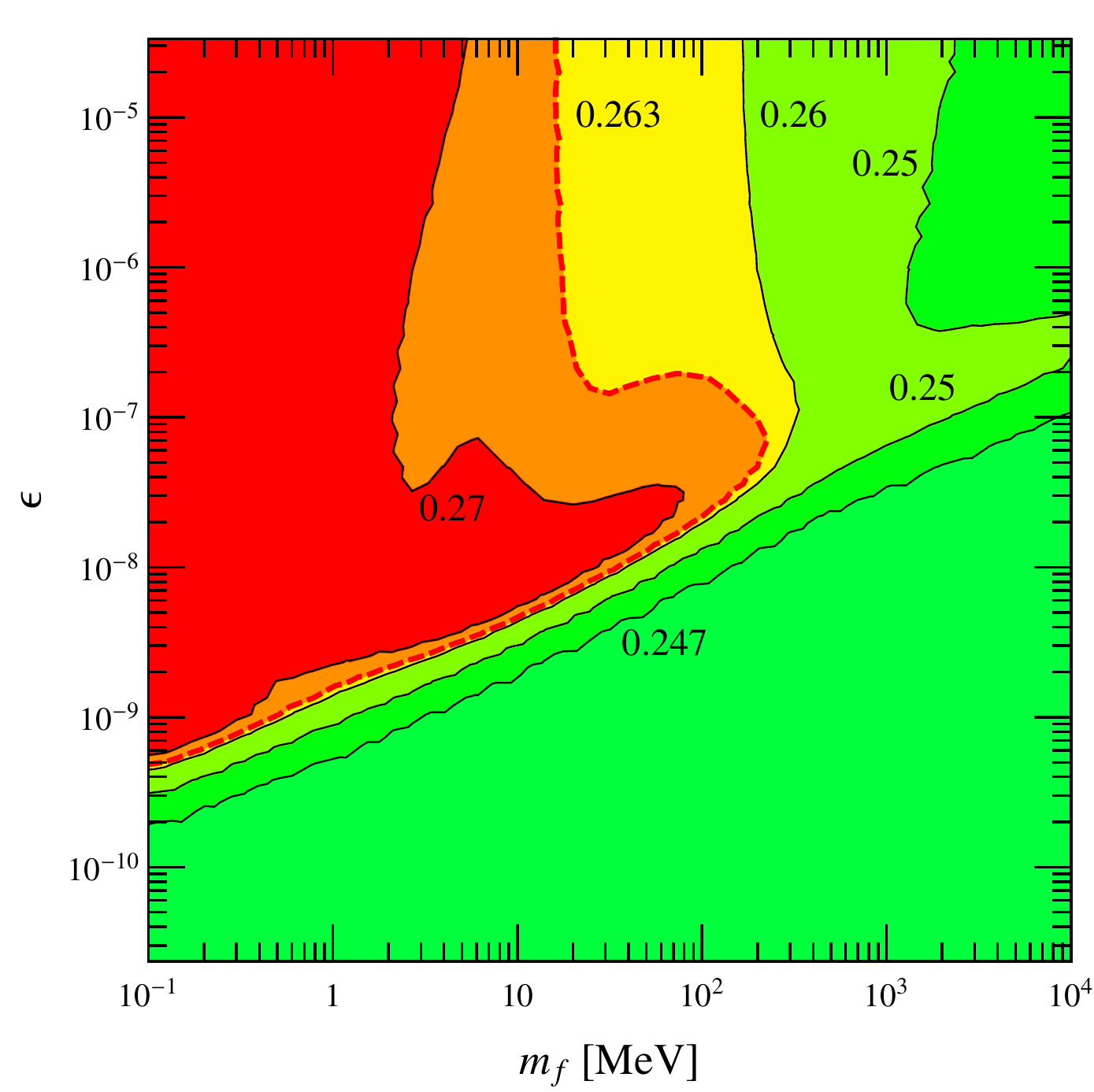}
\includegraphics[width=0.49\textwidth]{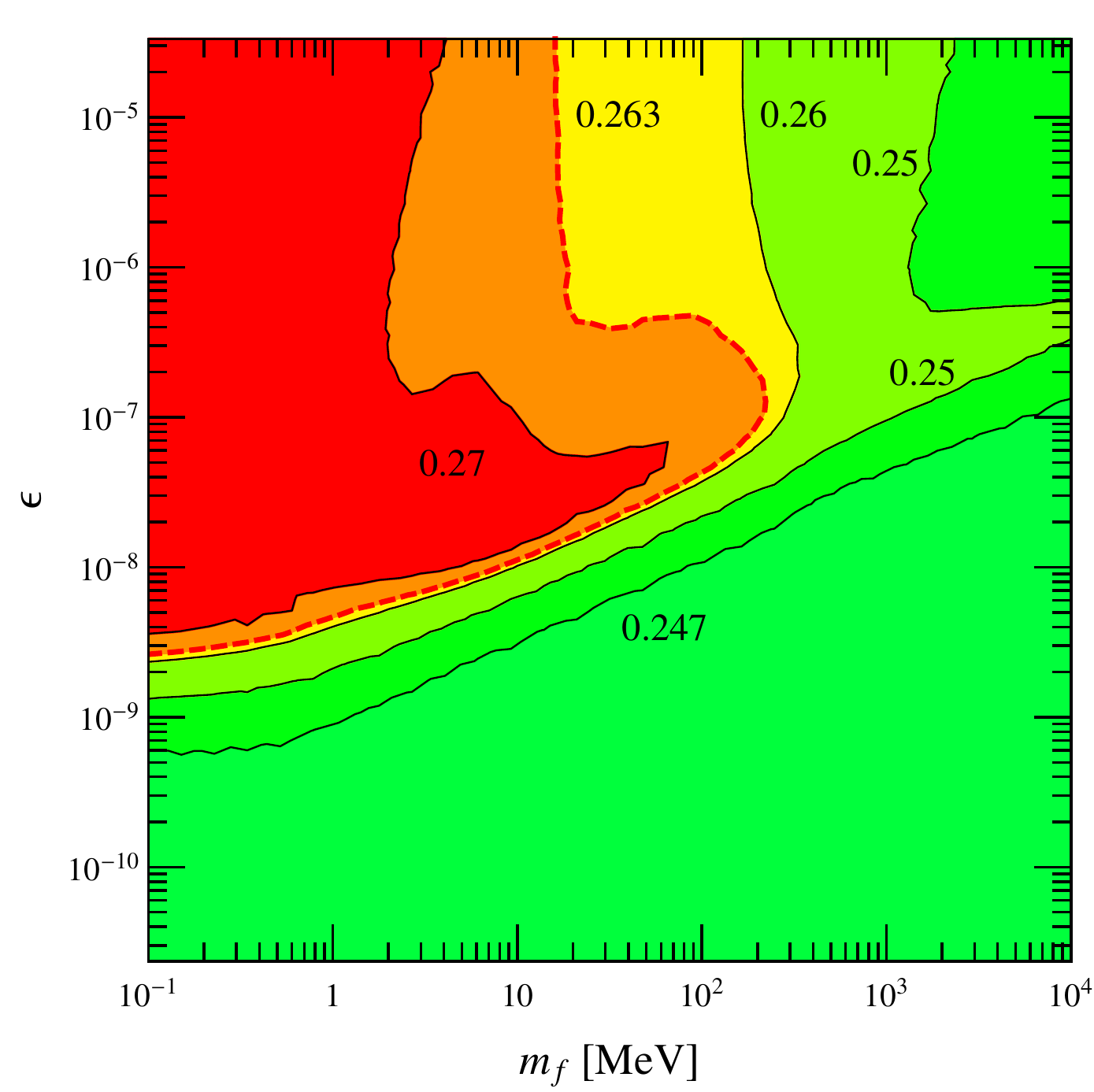}
\caption[Helium yield $Y_p$]{{\small{Helium yield $Y_p$ as a function of hidden fermion mass $m_f$ and minicharge $\epsilon$ for $g'=1$ (left) and $g'=0.01$ (right). Numbers correspond to the value of $Y_p$ of the closest contour line. Dark green coloring denotes regions far away from the upper limit $Y_p<0.263$ (95\% C.L.)~\cite{Mangano:2011ar}. The limit is given by the red dashed contour line. Orange and red regions are excluded by more than $2\sigma$ level.}}} \label{fig:heliumg1} \end{figure}

For $g'=1$ ($g'=10^{-2}$), the BBN results can be found in figure~\ref{fig:heliumg1}. 
Again, the contour lines for $g'=1$ are shifted towards smaller minicharges compared to $g'=0.1$ and the results for $10^{-2}$ are indistinguishable from $g'=0.1$.
 
\section{Conclusions}
 
In this paper we report our detailed calculation of the contribution of minicharged particles and hidden photons to the dark radiation of the Universe. Using our results we can conclude that the recent Planck data together with BBN constraints disfavors the existence of MCPs lighter than $\sim$ GeV unless their minicharge is very small $\lesssim 10^{-9}-10^{-7}$ (depending on the mass).  Our results extend to a broad range of hidden sector gauge couplings $g'\lesssim 0.1$ and we have also covered the case $g'=1$. 
The next generation of cosmological probes will be able to assess the existence of dark radiation with an estimated 1-$\sigma$ error of 0.05. Thus, we offer predictions of \neff  for a broad range of MCP masses and minicharges that we will allow in the future to strengthen the constraints or, more importantly to pinpoint the possible parameters of these elusive particles in case of a firm discovery of dark radiation.   
 
\section{Acknowledgements}

We would like to thank Yegor Goncharov, Thomas Hahn and Georg Raffelt for useful discussions. J.R. acknowledges support by the Alexander von Humboldt Foundation and partial support by the European Union through the Initial Training Network ÒInvisiblesÓ.
  
\appendix   
\section{$\neff$ from MCP decoupling in LTE \label{app1}}

If one assumes instantaneous decoupling, one can use entropy conservation to compute \neff for the case where a particle species decouples during the annihilation of another particle species. In our case, we compute \neff to be

\bea
\neff=&\,3\left(\frac{11}{4}\right)^{4/3}\left[ \frac{2}{2+7/2+g_{*S\text{DS}}(T^\dd_\nu)}\frac{2+g_{*Se^+e^-}(T^\dd_{\text{DS}})+g_{*S\text{DS}}(T^\dd_{\text{DS}})}{2+g_{*Se^+e^-}(T^\dd_{\text{DS}})} \right]^{4/3}\\+&\,\frac{8}{7} \left(\frac{11}{4}\right)^{4/3}\left[\frac{g_{*S\text{DS}}(T^\dd_{\text{DS}})}{2+g_{*Se^+e^-}(T^\dd_{\text{DS}})}\right]^{4/3},
\label{eq:app1}
\eea

where $(4/11)^{1/3}$ is the standard neutrino/photon temperature ratio,  $T^\dd_\nu$ ($T^\dd_{\text{DS}}$) is the decoupling temperature of the neutrinos (DS), $g_{*Se^+e^-}(T^\dd_{\text{DS}})$ are the entropy degrees of freedom of the electrons/positrons evaluated at the temperature of DS decoupling. 

\section{Astrophysical bounds at high masses}

The astrophysical bounds from red-giant, helium burning and white dwarf stars~\cite{Davidson:1991si,Davidson:1993sj,Davidson:2000hf,Raffelt:1996wa} are based on constraints on stellar energy loss. 
MCPs are produced by pairs in plasmon decay $\gamma^*\to f\bar f$ in stellar interiors and leave the star unimpeded contributing to the energy loss more efficiently than photons (only emitted from the surface). 
Transverse plasmons in such non-relativistic plasmas have a dispersion relation $\omega^2-k^2=\omega_\text{p}^2$.  
The relevant values for the plasma frequencies in the interior of helium-burning, Red Giant and white-dwarf stars are $\omega_\text{p}\sim 2, 18, 23$ keV~\cite{Raffelt:1996wa}. The energy loss per unit volume of transverse plasmon decay into massive MCPs is 
\be
Q=\left.\int_0^\infty \frac{k^2dk}{\pi^2} \frac{\omega \Gamma_{\gamma^*} }{e^{\omega/\TG}-1}
\right|_{\omega=\sqrt{\omega_\text{p}^2+k^2}}
\ee 
where $\TG$ is the plasma temperature and the plasmon decay rate into MCPs is given by 
eq.~\ref{plasmondecay}. We have denoted by $K^2=\omega^2-k^2$ the 4-momentum squared of the plasmon.  
The above equation can be straightforwardly extended into off-shell plasmons once we know their self energy $\Pi(\omega,k)$ in the medium because off-shell excitations are also thermally distributed~\cite{Weldon:1983jn}. Thus we have 
\be
Q=\int_0^\infty \frac{k^2dk}{\pi^2} \int_{2 m_f}^\infty\frac{\omega d \omega}{\pi}\frac{2 {\rm Im}\Pi}{(K^2-{\rm Re}\Pi)^2+({\rm Im}\Pi)^2}\frac{\omega \Gamma_{\gamma^*} }{e^{\omega/\TG}-1}
\ee 
In our case we can take $\Pi_T \simeq \omega^2_\text{p}+ i \omega \Gamma_T $. $\Gamma_T$ is the rate of Thomson scattering into free non-relativistic ambient electrons $\Gamma_T=n_e \sigma_T$ with $n_e$ the electron density and $\sigma_T=8\pi\alpha^2/3 m_e^2$ the Thomson cross section.   
Therefore we find
\be
\label{decau2}
Q=\int_0^\infty \frac{k^2dk}{\pi^2}\int_{2 m_f}^\infty \frac{d \omega^2}{\pi}\frac{\omega\Gamma_T}{(K^2-\omega_\text{p}^2)^2+(\omega\Gamma_T)^2}\frac{\omega \Gamma_{\gamma^*} }{e^{\omega/\TG}-1}
\ee 
The decay of the plasmon into the MCP pair requires $K^2>(2 m_f)^2$. When $\omega_\text{p}>2 m_f$ the 
new factor behaves like a delta function $d\omega^2 \delta(K^2-\omega_\text{p}^2)$ enforcing the  dispersion relation $\omega^2-k^2=\omega_\text{p}^2$ because typically $\Gamma_T\ll \omega_\text{p}$. 
The pole contribution dominates the $\omega-$integral and we recover the results of~\cite{Davidson:1991si,Davidson:1993sj,Davidson:2000hf,Raffelt:1996wa} which considered only on-shell plasmon decay.  
Contrarily, when $\omega_\text{p} <  2 m_f$ the pole does not contribute much and as we consider larger MCP masses soon becomes irrelevant. In this regime, our calculation reflects the process $\gamma+e^-\to e^- +f \bar f $ where the MCP pair is emitted through an off-shell photon after a common Thomson scattering of a thermal photon. We have neglected the contribution of electron-nucleus Bremsstrahlung $e+ Z\to e +Z +f  \bar f$ because it 
is subdominant at high MCP masses.    
In order to built the bounds shown in figure~\ref{fig:MCPresult} we have computed the integral \eqref{decau2} and match the constraint at low MCP masses with the already existing bounds~\cite{Davidson:1991si,Davidson:1993sj,Davidson:2000hf,Raffelt:1996wa}. We have colored as excluded all the regions where the millicharge $\epsilon$ is larger than the upper bound obtained. However, it is not completely clear what happens when we consider values of $\epsilon$ much above this boundary. For sure, the physics of stars will be very strongly modified but computing a self-consistent bound is extremely complicated. In principle there could be islands of parameter space where MCPs are trapped inside the star with a corresponding HP thermal bath and the energy loss is somehow quenched. 
We consider this unlikely, as a sizable amount of radiation would in any case be radiated for instance by our Sun and these particles should have produced some kind of signature on Earthly experiments.  We thus conclude that all the colored region is most likely excluded up to the largest values of $\epsilon$. 
Of course, these constraints are valid for models in which the minicharge arises by means other than the kinetic mixing. 

\section{Minicharged particles during recombination \label{CMBupdate}}

Following~\cite{Dubovsky:2003yn,Dolgov:2013una} we recomputed the bound on the MCP abundance during recombination. If MCPs couple strongly enough during recombination, they participate in the acoustic oscillations of the photon-baryon plasma. Comparing cosmologies with MCPs to the Planck data,~\cite{Dolgov:2013una} finds that the bound on the relic density of MCPs is
\be
\Omega_\text{MCP} h^2 < 0.001 \quad (95\% \ \text{CL}),
\ee
if the MCPs are strongly coupled to the plasma. This condition can be expressed as~\cite{Dolgov:2013una}
\be
\epsilon^2 \gtrsim 5\times 10^{-11} \text{GeV}^{-1/2} \frac{m_f}{\sqrt{\mu_{f,e}}+\sqrt{\mu_{f,p}}},
\ee
where $\mu$ is the reduced mass of the MCPs and electrons (protons), and the DS coupling $\alpha ' = g'^2/(4\pi) =0.1$ has been used. Integrating the usual Boltzmann equation for dark matter freeze out with $\alpha ' = 0.1$, we find an upper bound
\be
m_f< 241\ \text{GeV},
\ee
in the strongly coupled regime. This value agrees very well with the analytical prediction by~\cite{Davidson:1991si}. Figure~\ref{fig:MCPresult} shows this result.

\bibliographystyle{JHEP}
\bibliography{hiddenbiblio}

\end{document}